\documentclass[twocolumn,aps,prl,reprint,groupedaddress]{revtex4}
\usepackage{amsmath}
\usepackage{amssymb}

\usepackage{graphicx}
\usepackage{color}

\graphicspath{{images/}}

\usepackage{hyperref}
\hypersetup{
  colorlinks   = true, 
  urlcolor     = blue, 
  linkcolor    = blue, 
  citecolor   = blue 
}

\newcommand{\be}{\begin{equation}}
\newcommand{\ee}{\end{equation}}
\def\ba{\begin{aligned}}
\def\ea{\end{aligned}}

\newcommand{\bea}{\begin{eqnarray}}
\newcommand{\eea}{\end{eqnarray}}

\renewcommand{\Re}{{\rm \, Re\,}}
\renewcommand{\Im}{{\rm \, Im\,}}
\renewcommand{\vec}[1]{{\bf #1}}
\renewcommand{\hat}[1]{{\widehat #1}}

\renewcommand{\Im}{{\rm Im\,}}

\def\du{\,{\rm{div} \bf{u}}}

\begin{document}

\title{Magnetic field-induced giant enhancement of electron-phonon energy transfer
in strongly disordered conductors.}

\author{A. V. Shtyk $^1$, M. V. Feigel'man $^{1,2}$  and V. E. Kravtsov $^3$}
\affiliation{$^1$  L. D. Landau Institute for Theoretical Physics, Chernogolovka, Russia}
\affiliation{ $^2$ Moscow Institute for Physics and Technology, Moscow, Russia}
\affiliation{ $^3$ International Center for Theoretical Physics, Trieste, Italy}

\begin{abstract}
Relaxation of soft modes (e.g. charge density in gated semiconductor
heterostructures, spin density in the presence of magnetic field)
slowed down by disorder may lead to giant enhancement of energy
transfer (cooling power) between overheated electrons and phonons at
low bath temperature. We show that in strongly disordered systems
with time-reversal symmetry broken by external or intrinsic exchange
magnetic field the cooling power can be greatly enhanced. The
enhancement factor as large as $10^{2}$ at magnetic field $B \sim
10$ Tesla in 2D {\rm InSb} films is predicted. A similar enhancement
is found for the ultrasound attenuation.

\end{abstract}


\maketitle

{\it Introduction--} A number of recent experiments show that energy
transfer (the cooling power) $ \mathcal{J} (T_e,T_{ph}) =
J(T_{el})-J(T_{ph})$ between overheated electrons with temperature
$T_{el}
> T_{ph}$ and phonons at low bath temperature $T_{ph}$ may vary by
several orders of magnitude when measured per one electron per
volume. The out-flux $J(T)= W\, T^{p}$ may have different power-law
temperature dependence with the exponent $p$ both smaller and larger
than the classical result $p=5$ valid for pure metals. In disordered
metals with complete screening of Coulomb interaction and impurities
that are fully involved in the lattice motion one expects
~\cite{schmid, reyzer,kravtsov_yudson} a power law with $p=6$ which
corresponds to weaker energy transfer compared to the clean case.
This is related with the ''Pippard ineffectiveness condition''~
(denoted as PIC below) \cite{pippard,akhiezer} formulated for the
rate of inelastic electron-phonon scattering. A very accurate
experiments in metal films of Hf and Ti~\cite{Gersh2001} confirmed
this theoretical expectation, including the value of the pre-factor
$W$ in front of $T^{6}$. At the same time, experiments on heavily
doped {\rm Si}\cite{Pekola2004} which also demonstrated the $T^{6}$
behavior, gave at low temperatures the value $W/n_{e}$ ($n_{e}$ is
the carrier density) larger by a factor of $10^{3}$ than the
theoretical prediction in Ref.\cite{schmid, reyzer,kravtsov_yudson}.
Surprisingly, the $T^{6}$ behavior of the cooling rate with
approximately the same values of $W/n_{e}$ as in
Ref.\cite{Pekola2004} were extracted from the recent experiments
\cite{Sacepe2009} on amorphous {\rm InO} films showing {\it weakly
insulating} behavior in magnetic field of 11T. In this case
$W/n_{e}$ was larger by a factor of $5\times 10^{4}$ than the
theoretical prediction for a dirty metal approaching  the Anderson
transition. Clearly, neither of the above cases with anomalously
large cooling rate correspond to the piezoelectric type of
electron-phonon coupling where the PIC does not hold and the theory
predicts $T^{4}$ temperature behavior of cooling rate \cite{piezo,
Gersh2000}. It is also dubious that the model of impurities which
are only partially involved in the lattice motion \cite{reyzer} that
also leads to enhanced cooling rate with $T^{4}$ temperature
behavior, is realistic for the cases in question. Thus there was a
quest from experiment for a different and more general mechanism of
enhancement of cooling rate in strongly disordered conductors.

In the present Letter we demonstrate existence of a general
mechanism which is capable of enhancing by a factor $10^{2}-10^{3}$
of both the cooling power $J(T)$ and ultrasound attenuation
$\tau^{-1}_{ph}$ (for longitudinal phonons) at low temperatures.
This mechanism is effective if lattice motion is able to induce
significant oscillations of {\it local} densities of certain
globally conserved physical quantities. The deviations of these
local densities from their equilibrium values are enhanced by slow
diffusive character of electron motion (characterized by both small
frequency and small momentum) aimed to restore equilibrium. This
leads to a significant retardation  in the response and thus to the
entropy production and dissipation. The proposed mechanism is
reminiscent of the Mandelstam-Leontovich (ML) mechanism of phonon
attenuation in liquids \cite{MLref}.  In contrast to PIC which
suppresses the relaxation rate $\tau_{ph}^{-1}$ at strong disorder
and small carrier concentration, the ML mechanism is efficient at
these conditions.
The particular realizations of such a mechanism were studied
previously in the literature. Specifically, relatively weak Coulomb
interaction between electrons in semiconductors, when  the local
 electroneutrality condition
 is not strictly obeyed and the density fluctuations are not completely suppressed,
 was a cause of enhancement of cooling rate
 discussed in Ref.\cite{Reizer_screening}.
The asymmetric inter-valley modes  were shown \cite{Prunnila} to
lead to a significant enhancement of cooling rate in multiple-valley
semiconductors such as ${\rm Si}$. Below we reproduce some of these
results from our general approach.

However, a really new effect we are predicting is the giant
enhancement of the cooling rate and ultra-sound attenuation in the
presence of external magnetic field or in ferromagnetic materials
where the role of external magnetic field is played by the intrinsic
exchange field. In this case it is the spin-density mode which can
be excited by absorption of a phonon or terminated by creation of a
phonon, that is responsible for the enhanced ultrasound attenuation
or the enhanced cooling rate.

{\it Cooling power and ultrasonic attenuation--} The starting point
of our consideration is the quantum kinetic equation
\cite{SerMit,apps} for the phonon distribution function
$B_{ph}(\omega,T)$. for the case of partial equilibrium in the
electronic (with the temperature $T_{el}$) and phonon (with the
temperature $T_{ph}$) systems at the lack of the total equilibrium
($T_{el}> T_{ph})$:
\be
	\partial_tB_{ph}(\omega,T_{ph})
	=\left[B_{ph}(\omega,T_{el})-B_{ph}(\omega,T_{ph})\right]\cdot\tau^{-1}_{ph}\left(\omega,T_{el}\right).
\ee
If the electron-phonon energy relaxation is much slower than the
electron-electron one and the phonon system is well coupled to the
thermostat (fridge), a quasi-equilibrium situation with two
temperatures is realized. In this approximation
$B_{ph}(\omega,T)\equiv B_{ph}(\omega/T)=
\frac12(\coth(\omega/2T)-1)$ is the equilibrium phonon distribution
function. The phonon decay rate $\tau^{-1}_{ph}$ is then given by
the imaginary part of the phonon self-energy
$\Sigma^{R}(\omega,q,T)$:
\be \label{tau-phon}
	\tau^{-1}_{ph}(\omega,T_{el})=\frac{1}{\rho_{m}\,\omega}\,{\rm Im}
	\Sigma^{R}(\omega,q,T_{el})|_{\omega=v_{s}q}. 
\ee
The phonon decay rate depends only on the electron temperature,
since the (weak) electron-phonon interaction is considered in the
leading approximation, and thus the phonon self energy (which is
second order in the e-ph coupling) is expressed in terms of
electronic variables only. If in addition, the effect of
electron-electron interaction is reduced to charge screening
considered in the RPA approximation, the phonon relaxation rate
$\tau^{-1}_{ph}(\omega,T_{el})\equiv \tau^{-1}_{ph}(\omega)$ does
not depend explicitly on the electron temperature.

Now the energy flow    $\mathcal{J} = \frac{dE_{ph}}{dt}$ from hot
electrons to cool phonons can be written as follows
$\mathcal{J}=J(T_{el})-J(T_{ph})$, where:
\be 
	\label{eq:heat_flow} J(T) =
	\int\limits_0^{\infty}d\omega\,\omega\,
	  \nu_{ph}(\omega)\,
	\frac{B_{ph}(\omega/T)}{\tau_{ph}(\omega)} 
\ee
and $\nu_{ph}(\omega)=\omega^2/(2\pi^{2}v_{s}^{3})$ is the phonon
density of states for 3D phonons with the sound velocity $v_{s}$.
Eqs.(\ref{tau-phon}),(\ref{eq:heat_flow}) establish a relationship
between the cooling rate $J(T)$ and the attenuation time
$\tau_{ph}(\omega)$ of ultrasound with the frequency $\omega$.   In
particular, it follows from Eq.(\ref{eq:heat_flow}) that for the
power-law dependence $\tau^{-1}_{ph}(\omega)\propto \omega^{\beta}$,
the cooling rate due to 3D phonons is proportional to $J(T)\propto
T^{4+\beta}$.

{\it Local and diffusion contribution to cooling rate}--  In impure
conductors there are two distinctly different contributions to the
phonon relaxation time. One is local and determined by the small
distances $|{\bf r}-{\bf r'}|\sim l$ between the points ${\bf r}$
and ${\bf r'}$ of phonon absorption  and re-emission. The other one
allows many scattering events of electrons off impurities between
the points ${\bf r}$ and ${\bf r'}$. This is the diffusive
contribution. With increasing disorder and decreasing the mean free
path $l$ the local contribution diminishes. This leads to the so
called {\it Pippard inefficiency condition} (PIC) when the
relaxation rate   $\tau^{-1}_{ph}(\omega)$ of phonons with momentum
$q$ is proportional to $lq^{2}\ll q$ instead of
$\tau_{ph}^{-1}\propto q$ for longitudinal phonons in the clean case
\cite{pippard,akhiezer}:
\be 
	 \frac{1}{\tau^{(PIC)}_{ph}(\omega)}= c_\alpha
	  \frac{2\nu
	p_F^2}{\rho_m}Dq^2 \sim \frac{m\,n_{e}}{\rho_{m}}\,Dq^{2},
	\label{eq:tau_cl} 
\ee
where $\nu$ is an electron density of states per spin\cite{nu}, $p_F
= m v_{F}$ is Fermi momentum, $n_{e}$ and $\rho_m$ are the density
of electrons and the mass density of material, $D = v_F l/d_e$ is
the diffusion coefficient, $q=\omega/v_{s}\ll 1/l$ is the phonon
momentum, $d_e$ is the dimensionality of electron motion. The
subscript $\alpha$ corresponds to the choice of either transverse
(tr) or longitudinal (l) phonons; correspondingly, numerical
coefficients $c_\alpha$ are defined as $c_{tr} =1/(2+d_e)$ and $c_l
= 2(1-d_e^{-1})/(2+d_e)$.

The diffusion contribution has an opposite trend and increases with
increasing disorder. The goal of our paper is to analyze this very
contribution in different physical situations.

We will use the co-moving frame of reference (CFR) bound to the
lattice and impurities rigidly imbedded in it and moving in the
laboratory frame of reference (LFR). Then for a single branch of
electrons, one finds \cite{blount,schmid} for electron-phonon
interaction in the CFR:
\be \label{eq:e-ph_interaction} H_{e-ph}=
-\sum_{{\bf p},{\bf q}}p_\alpha (v_\beta\nabla_{\beta} u_\alpha
)_{q} \cdot \bar\psi_{p} \psi_{p+q} \ee
where $\vec p$ and $\vec v$ denote  electron momentum and velocity,
respectively and $\vec u$ is the lattice displacement. Note that
this term appears due to the inhomogeneous Galilean shift ${\bf
p}\,d{\bf u}({\bf r}(t),t)/dt$ of the energy of a quasi-particle at
a point ${\bf r}={\bf r}(t)$, while the usual deformation potential
in LFR is canceled by the modification of ${\rm e-e}$ interaction
due to inhomogeneous coordinate transformation \cite{tsuneto}
$\widehat{H}\rightarrow \widehat{U}^{-1}\widehat{H}\widehat{U}$ with
$\widehat{U}=1+\frac{1}{2}\{ {\bf u},\boldsymbol{\nabla}\}$. The tensor structure
of Eq.(\ref{eq:e-ph_interaction}) is crucial for local processes
only, while for diffusion processes, it is sufficient to average the
e-ph vertex over the Fermi surface. For a metal with isotropic
electron dispersion one finds $\Gamma=\overline{p_\alpha
(v_\beta\nabla_{\beta} u_\alpha )_{q}}=p v/d_e\,{\rm div}{\bf u}$.
In general $\Gamma$ may contain other contributions. In particular,
for semiconductors $\Gamma$ is known~\cite{dp_semiconductors} to be
much larger than $E_F$ due to contribution $\Gamma_{bs}$ originating
from the shift of conduction band-edge $E_{b}$.

Under the condition of strict electroneutrality, the scalar vertex
$\Gamma$ is screened out completely and  the classical result
Eq.(\ref{eq:tau_cl}) is valid. This is not the case, however, when
$N$ different types of quasiparticles are
 present \cite{Prunnila}.
Then the interaction can be written as
\be \label{new-e-ph} H_{e-ph} = \sum_{j=1}^N \Gamma^{(j)}\,{\rm
div}{\bf u}\; (\bar\psi\psi)_j , \ee
where ($\nu_{j}$ is the partial DoS at the Fermi level):
\be\label{Gamma-band} \Gamma^{(j)}= -
p^{(j)}v^{(j)}/d_{e}+\Gamma_{bs}.\ee
Note that Eq.(\ref{new-e-ph}) is principally different from the
${\rm e-ph}$ interaction in the LFR, even when $\Gamma_{bs}=0$. The
latter contains the deformation potential
$\Gamma_{def}=\sum_{j}\nu_{j}\,(p_{F}^{(j)}v_{F}^{(j)}/d_{e})/\sum_{j}\nu_{j}$
which is {\it symmetric} in the electron branch indices $j$, as well
as the moving-impurity part \cite{schmid, reyzer, kravtsov_yudson}.
The latter part leads to the mode-asymmetry of ${\rm e-ph}$
interaction in LFR which in CFR is provided by the Galilean shift
term.

 The Coulomb
interaction is able to screen out only the single mode corresponding
to the total density $n= \sum_j (\bar\psi\psi)_j$, whereas $N-1$
asymmetric modes survive screening~\cite{Prunnila,blount}. Their
slow, diffusive character in strongly disordered conductors may lead
to a considerable enhancement of the cooling rate and ultrasound
attenuation.  The particular case of the effect of such unscreened
diffusion modes was studied in Ref.\cite{Prunnila} for the case of
$N$ species of electrons corresponding to N inequivalent valleys in
semiconductors.

Below we present a simple derivation of the diffusion-enhanced
contribution to the phonon relaxation rate $\tau^{-1}_{ph}$ in terms
of macroscopic equations for the current and density of electrons;
alternative diagrammatic derivation is presented in
\cite{supplementary}, Sec. III. In the CFR the continuity and
diffusion equations for each i-th species of quasiparticles read:
\be
\label{eq:equation_motion}
\left\{
\begin{aligned}
&\partial_t n^{(i)}+\mathrm{div} \vec j^{(i)}=0,
\\
&\vec j^{(i)} =-D\vec \nabla n^{(i)} - \kappa_{i} \vec F^{(i)} 
\ea \right. \ee
where $(i)$ stands for the quasiparticle branch number, $n^{(i)}$ is
the electron density, $\vec j^{(i)}$ is the particle number current,
$\kappa_i = \nu_i D_i$ is the mobility, $D_i$ is the diffusion
coefficient for the $i$-th branch,
 $\vec F^{(i)}=-\nabla U^{(i)}$ and $U^{(i)}$ is the potential energy.
In the simplest derivation we assume no inter-branch mixing and thus
the continuity equations in Eq.(\ref{eq:equation_motion}) imply that
each of the partial electron densities $n^{(i)}$ are conserved
separately. Generalization to the case where there is mixing between
the branches will be done at the end of the paper. The potential
energy $U^{(i)} = U_C+\Phi^{(i)} $ in Eq.(\ref{eq:equation_motion})
consists of the Coulomb part $U_C$ and the phonon part $\Phi^{(i)} =
\Gamma^{(i)}\du$:
\be
U^{(i)}= \int {\cal V}_0(\vec r-\vec r')\sum\limits_{j}\delta n^{(j)}(\vec r') +
\Gamma^{(i)}\du
\label{potential}
\ee
where ${\cal V}_0(\vec r)$ is the bare Coulomb potential acting
between conduction electrons; below we use its Fourier-transform
$V_0(q)$. Note that $V_0(q)$ does not include screening by
conduction electrons in the sample.

Eqs.(\ref{eq:equation_motion}),(\ref{potential}) is a full set of
equations describing the diffusion and screening of partial
densities $n_{i}$. Let us first study their solution in the case of
perfect screening and multiple electron branches ($N>1$). It
formally corresponds to $\Pi(\omega,q)\,V_{0}(q)\gg 1$, where
$\Pi(\omega,q)$ is the total polarization function. For the density
modulation $n^{(i)} (\omega,q)$ induced by the phonon with frequency
$\omega$ and momentum $q$ one finds from
Eqs.(\ref{eq:equation_motion},\ref{potential}):
\be n^{(i)}(\omega,q) = \Pi_i (\omega,q) \left(\Phi_i(\omega,q) -
\Phi_C (\omega,q)\right) \label{solution-n} \ee
where  $\Phi_i (\omega,q) = \Gamma^{(i)} \du $ , $
\Phi_C={\sum_j\Phi_j\Pi_j}/\sum_{j}\Pi_j$ represents dynamical
screening of Coulomb interaction and
$\Pi_i=\nu_iD_iq^2/(-i\omega+D_iq^2)$ is the partial polarization
function. The solution Eq.(\ref{solution-n}) obeys
charge-neutrality: $n_{tot}=\sum n^{(i)}=0$.

The diffusion contribution to the phonon decay rate  may be
expressed as $\tau_{ph,ML}^{-1}=\frac{|Q_t|}{E_w}$, where $Q_{t}$
and $E_{w}$ are the dissipation power and the acoustic wave energy
in a unit volume, respectively:
\be \label{eq:Q,E} Q_t=\frac{1}{2}\Re\left(\vec j^*\cdot \vec F
\right)\, , \qquad E_w= \frac{\rho_m}{2}\omega^2u_m^2. \ee
Here $u_m$ is an amplitude of ionic displacement  and $\vec u=(\vec
q/q)u_m\exp[-i\omega t+i\vec q \cdot\vec r]$. Below we apply
Eq.(\ref{eq:Q,E}) to compute $\tau_{ph}^{-1}(\omega)$.

{\it Giant enhancement by magnetic field-- } The case of $N=2$
quasiparticle branches has a very important application. It
corresponds to the two spin projections. However, they should be
inequivalent with respect to the coupling to phonons. This is a
consequence of the general statement that the spin density can be
excited by phonons only if time-reversal invariance (TRI) is broken.
First, we discuss the case when TRI is broken by external magnetic
field.

For $N=2$ a simple calculation based on Eqs.
(\ref{solution-n}),(\ref{eq:Q,E}) leads to the following expression
for the diffusion contribution to the decay rate of acoustic phonon:
\be \label{eq:tau_mb} \tau_{ph}^{-1}(q)=
\frac{\left(\Gamma_1-\Gamma_2\right)^2}{\rho_m}\frac{\nu_* D_*
q^2}{v_{s}^2 + D_*^2q^2} \ee
where $v_{s} = \omega/q$ is the sound velocity, while
$\nu_*=(\nu_1^{-1}+\nu_2^{-1})^{-1}$ and
$D_*=\nu_*^{-1}((\nu_1D_1)^{-1}+(\nu_2D_2)^{-1})^{-1}$ are the effective density of states
and diffusion coefficient, respectively.

When a parallel magnetic field $H$ is applied   to a two-dimensional
electron gas   the bottom of the spin-down and spin-up conduction
bands get shifted by $ \pm (1/2)\mu H$ with respect to their
position at $H=0$. This leads to a change of $\delta(p_{F}v_{F})=\pm
\mu H$, where $(1/2)\mu = (g/2) \mu_B$ is the electron magnetic
moment. Thus from Eq.(\ref{Gamma-band}) we conclude that an
asymmetry $\Gamma_\uparrow-\Gamma_\downarrow=(2/d_{e})\,\mu H$,
arises due to the Galilean shift of the quasi-particle energy.
Then, according to Eq.(\ref{eq:tau_mb}), the phonon relaxation rate
acquires an $H$-dependent contribution that may become dominant at
sufficiently strong field and low phonon frequencies. Adding the
local contribution (\ref{eq:tau_cl}) and the
magnetic-field-controlled diffusion contribution,
Eq.(\ref{eq:tau_mb}), one finds
 for the full phonon decay rate
 $\tau_{ph}^{-1}= [\tau^{(PIC)}_{ph}]^{-1} \cdot \mathcal{F}_{H}(q,h)$,
 where for ${\bf q}$ parallel to 2D gas:
\be \mathcal{F}_{H}(q,h) =  1+\frac{v_F^2h^2}{v_{s}^2 + (Dq)^2}.
\label{enchance} \ee
Here $\tau^{(PIC)}_{ph}$ is given by Eq.(\ref{eq:tau_cl}) for
longitudinal phonons and $h=(|\mu| H/2\varepsilon_F)$ \, (we assume
here $h\ll 1$).
 The
enhancement factor $\mathcal{F}_{H}$
 can become very large for strong spin polarization, $ h \sim 1$. In particular, for  low
phonon momentum,  $ ql \leq  v_{s}/v_F$,   the factor
$\mathcal{F}_{H}$ is of the order of inverse adiabatic parameter
 $(v_F/v_{s})^2 \sim 10^5$.
The strong spin-orbit interaction which leads to mixing of spin-up
and spin-down branches sets limitation on the enhancement factor.
Its maximum value becomes ${\cal F}_{max}\sim \tau_{SO}/\tau$
($\tau_{SO}$ is the spin relaxation time and $\tau\ll \tau_{SO}$ is
the momentum relaxation time) instead of ${\cal F}_{m}\sim
(v_{F}/v_{s})^{2}$. This makes the optimization of parameters to
maximize the enhancement factor a hard problem, since materials with
large $g$-factor (to maximize $\mu H$) usually have large spin-orbit
coupling. Nevertheless the example of ${\rm InSb}$ films shows that
${\cal F}\sim 10^{2}$ is experimentally achievable (see Fig.1).

{\it Enhancement of cooling rate in ferromagnetic metals--} Another
relevant example is
 provided by ferromagnetic metals with strong intrinsic band-splitting due to the exchange field.
  In the case of  $\mathrm{Fe}$: $\mu H^*\approx1.8\,\mathrm{eV} ,\,\varepsilon_F=11.1\,
  \mathrm{eV},\,v_F=1.98\times10^8\mathrm{cm/s},\, v_{s}\approx6\times10^5\mathrm{cm/s}$. The spin relaxation rate may be estimated as $\tau/\tau_{SO}\sim (\alpha Z)^4\sim10^{-3}$, $\alpha=1/137$ and $Z=26$ being the fine structure constant and the atomic number respectively,
  resulting in the maximum enhancement of the phonon relaxation time as large as
  $\mathcal{F}_{H}\sim(\mu H^*/\varepsilon_F)^2\,(\tau_{SO}/\tau)\sim 10$.

{\it Enhancement by incomplete screening.} In the case of a single
quasipaticle branch, the general approach
Eqs.(\ref{eq:equation_motion}),(\ref{potential}),(\ref{eq:Q,E})
describes the diffusion-enhanced dissipation due to violation of the
 charge neutrality condition at a large screening length.
In this case we obtain $n=2\nu D q^2
\left[-i\omega+Dq^2+2\kappa\,q^2\,V_0(q)\right]^{-1}\,\Phi(\omega,q)$
and the enhancement factor:
\be \label{eq:F_C} \mathcal{F}_{C}= 1 +
\frac{c_l^{-1}(\Gamma/p_Fv_F)^2}{(v_{s}/v_F)^2+d_{e}^{-2}(q^2l^2)(1+2\nu
V_0(q))^2} \ee
For the 2D gas with Coulomb interaction and the constant dielectric
permittivity $\varepsilon$ of surrounding media we have
$V_0(q)=V_{2D}(q)=2\pi e^2/\varepsilon q$. In the relevant range of
${\bf q}$ parallel to the 2D gas the $\mathcal{F}_C(q)$ factor
reduces to a constant.
This corresponds   to the cooling rate $J(T)\propto T^{6}$
\cite{Reizer_screening} but with the enhanced pre-factor
proportional to $\varepsilon^{2}/g_\Box^2$ (where $g_\Box$ is
dimensionless conductance per square in $e^2/h$ units) at strong
disorder and large dielectric constant $\varepsilon$. In 3D
conductors, when $V_{0}(q)\propto q^{-2}$ Eq.(\ref{eq:F_C}) has a
regime were ${\cal F}_{C}\propto q^{2}$. Correspondingly, the
cooling rate appears   to be $J(T)\propto T^{8}$
\cite{Reizer_screening}.

An interesting situation arises in 2D electron gas in the presence
of a gate that additionally screens Coulomb interaction and allows
the density to fluctuate stronger. For this geometry and ${\bf q}$
parallel to 2D gas $V_{0}(q)\rightarrow
V_g(q)=V_{2D}(q)\,(1-e^{-2qb})$, $b$ being the distance between the
2d electron gas and the gate. For phonons with the wavelengths $1/q
\geq b$, the effective potential $V_g(q)\approx V(q)\cdot 2qb\approx
const$ and the presence of adiabatic parameter in the denominator of
(\ref{eq:F_C}) does become important at low enough temperatures:
\be \label{eq:tau_C_gate} \mathcal{F}_{C_{gate}}= 1 + \frac{4
(\Gamma/p_Fv_F)^2}{(v_{s}/v_F)^2+(q^2l^2)(4\pi\nu
e^2b/\varepsilon)^2}, \ee
where Coulomb interaction is still assumed to be relatively strong:
$2\pi\nu e^2b/\varepsilon\gg1$. In this case there is a regime where
the enhancement factor is proportional to $q^{-2}$, and the cooling
rate $J(T)\propto T^{4}$.

{\it Modes mixing and realistic example--} Finally, we collect
results of both the ML enhancement due to the charge density and the
spin density fluctuations, taking also into account mixing of spin
projections by the spin-orbit interaction characterized by the
parameter $\tau_{so}^{-1}$. We also consider the dependence of
relaxation rate on the direction of phonon propagation relative to
2D gas\cite{supplementary}. Both effects lead to the replacement $
Dq^2/(-i\omega + Dq^2) \Rightarrow D{\bf q}_{\parallel}^2/(-i\omega
+ D{\bf q}_{\parallel}^2 + 1/2\tau_{so})$. It results in the total
enhancement factor of the form:
\be {\cal F} ={\cal F}_{C_{2D}}+\frac{{\bf q}^{2}_{\parallel}
v_F^2h^2}{(q v_{s})^2 + (D{\bf
q}_{\parallel}^{2}+1/2\tau_{so})^2}
 \label{eq:F_so}
\ee
For 2D electrons and 3D phonons $|{\bf q}_{\parallel}|=q\,\sin \theta$
is the phonon momentum component parallel to the 2D system which
appears in all the terms originating from electron diffusion. In
this case ${\cal F}_{C_{2D}}$ is independent of $q$, and  ${\cal F}$
has a maximum as a function of $\omega$. The spin fluctuation effect
given by the second term vanishes at small $\omega$ because of the
mixing of branches caused by spin-orbit interaction. It also
decreases at large $\omega$ because the dissipation power increases
slower with $\omega$ than does the acoustic wave energy.
At large
enough Zeeman splitting $h$ when the effect of spin fluctuations in
its maximum is large, there is a wide frequency  region (the falling
part of the curve ${\cal F}\propto \omega^{-2}$ in Fig.1)
where $\tau_{ph}^{-1}$ is almost frequency independent. In this region   the cooling/heating rate $J(T)\propto
T^{4}\,\ln T$ for the quasi-2D case.
This temperature dependence is almost the same as in the case of
impurities which are not fully involved in the lattice motion~\cite{static_disorder}.
 The
extra logarithmic factor arises because of the angular averaging of
$1/\tau_{ph}(\theta)$ dominated by the small values of $\theta$.
To illustrate this behavior we consider a thin film of semiconductor
$\mathrm{InSb}$ (g-factor $|g|\approx50$).  At strong (and parallel to
the 2D plane) magnetic fields $|g|\mu_{B}H\gg \Delta_{SO}$
classification in terms of the spin subbands is still valid
approximately, in spite of the Rashba spin-orbit coupling
$\Delta_{SO}$. The analysis presented in~\cite{supplementary}, Sec. IV, VI,
leads to Eq.(\ref{eq:F_so}) and is summarized in Fig.1.
\begin{figure}[h]
\center{\includegraphics[width=1\linewidth]{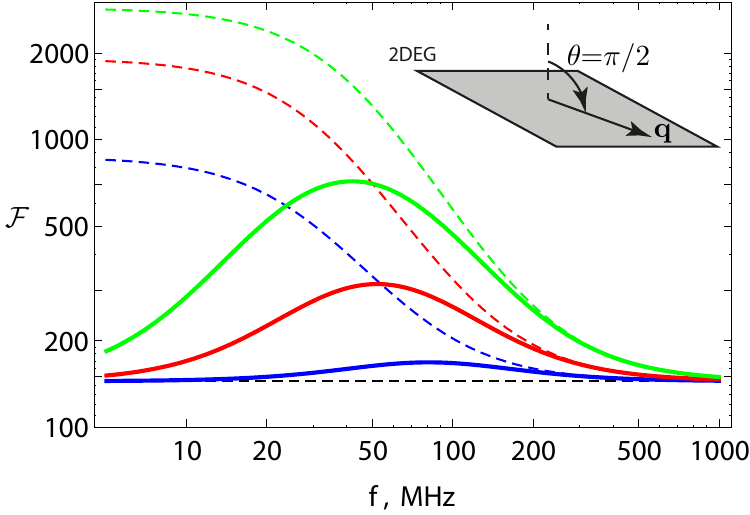}} \label{fig:1}
\caption{(Color online) The total enhancement factor
$\mathcal{F}=\tau^{(0)}_{ph}/\tau_{ph}$ at $\theta=\pi/2$ of
ultrasound attenuation in the 2D semiconductor InSb. The parameters
 taken are $n=10^{11}cm^{-2}$, $p_Fl=50$, $\Delta_{SO}=0.1meV$ and magnetic fields are $3\,\mathrm{T}$ (blue), $5\,\mathrm{T}$ (red)
and $7\,\mathrm{T}$ (green). Dashed curves represent the result in the absence of SO relaxation, $\Delta_{SO}=0$. The black dashed curve gives the enhancement  $\mathcal{F}_C$ by incomplete screening.}
\end{figure}

In conclusion, we  demonstrated an existence of a general relaxation
mechanism that leads to enhancement of both the e-ph cooling power
and the phonon decay rate.  In particular, it may lead to a strong
enhancement of the cooling power in disordered conductors in the
external magnetic field or in disordered ferromagnetic metals.

We are grateful to S. Dorozhkin, M. Gershenson, J. Pekola, M. Reznikov and K. Tikhonov for useful discussions.
The research done by  M.V.F. and A.V.S was supported by the RFBR grant \# 10-02-00554.
A.V.S. also acknowledges support from Dynasty foundation.


\onecolumngrid

\clearpage
\twocolumngrid

\renewcommand{\theequation}{S\arabic{equation}}
\renewcommand{\thefigure}{S\arabic{figure}}

\centerline{\large\textbf{Supplementary Material}}

\vskip 12mm

\section{I. Hamiltonian}

We consider an electron system with several spectral branches (which
are labeled by $ i \in (1,N)$ ).  Interaction between electrons is
considered in the direct (density-density) channel only, and will be
treated within the random phase approximation (RPA).  Electrons also
scatter off local impurities, and we assume this scattering to be
identical for all spectral branches, $U_{imp,i}(\vec r)=U_{imp}(\vec
r)$. While treating electron-phonon inelastic processes, it is
convenient to work in the co-moving frame of reference (CFR), where
the effective Hamiltonian acquires the following
form~\cite{s_schmid,s_reyzer,s_kravtsov_yudson} in the momentum
representation:
\bea
\label{seq:H}
&&H=H_{e0}+H_{e-e}+H_{e-ph},
\\\notag
&&H_{e0}=\sum_{\vec p, i}\xi_i(\vec p)\overline{\psi}_{\vec p,i}\psi_{\vec p,i}+
\sum_{\vec p,\vec p',i}U_{imp}(\vec p'-\vec p)\overline{\psi}_{\vec p,i}\psi_{\vec p',i},
\\\notag
&&H_{e-e}=\frac{1}{2}\sum_{\vec q} V_0(\vec q)n_{\vec q}n_{-\vec q},
\\\notag
&&H_{e-ph}=\sum_{\vec q,i} \Gamma_i(\du)_{\vec q} n_{-\vec q,i},
\eea
where $U_{imp}(\vec r)$ is disorder potential, $V_0(q)$ is a bare
Coulomb interaction attenuated by the background dielectric constant
$\varepsilon$ which could be additionally screened by a metallic
gate;  $n_{\vec q, i}=\sum_{\vec p}\overline{\psi}_{\vec p +\vec
q,i}\psi_{\vec p,i}$ and $n_{\vec q}=\sum_i n_{\vec q,i}$ are the
i-th branch partial- and total densities, respectively. $\Gamma_i$
and $(\du)_{\vec q}=i\vec q\cdot \vec u$ are the e-ph interaction
constant averaged over the Fermi surface and the divergence of ionic
displacement field $\vec u$. The Hamiltonian (\ref{seq:H}) contains
no inter-branch scattering. We assume these processes to be much
weaker than the intra-branch scattering and discuss their role later
on in Sec.III C.

For convenience, we present an explicit derivation of Hamiltonian in CFR.

\subsection{Hamiltonian in LFR}

In LFR electron-phonon interaction is mediated via two processes \cite{s_kravtsov_yudson}. First, electron-phonon-ion, when phonons disturb positive ionic background density, $n_{ion}(\vec u)=n_{ion}(1-\du)$. Second, when phonons displace impurities, $\vec r_{imp}(\vec u)\rightarrow \vec r_{imp} +\vec u(\vec r_{imp})$, and thus affect random potential $U(\vec r)\rightarrow U(\vec r) - \nabla_\alpha(u_\alpha(\vec r) U(\vec r))$:
\bea
\label{seq:H_LFR}
&&H^{LFR}=H_{K}+H_{U}+H_{e-e}+H_{e-ph}^{LFR},
\\\notag
&&H_{K}=\sum_{\vec p, i}\xi_i(\vec p)\overline{\psi}_{\vec p,i}\psi_{\vec p,i},
\\
\notag
&&H_{U}=\sum_{\vec p,\vec p',i}U_{imp}(\vec p'-\vec p)\overline{\psi}_{\vec p,i}\psi_{\vec p',i},
\\\notag
&&H_{e-e}=\frac{1}{2}\sum_{\vec q} V_0(\vec q)n_{\vec q}n_{-\vec q},
\\\notag
&&H_{e-ph}^{LFR}=H_{e-ph-ion}^{LFR}+H_{e-ph-imp}^{LFR},
\\
&&H_{e-ph-ion}^{LFR}=\sum_{\vec q} \left[(n_{ion}\du)_{\vec q} V_0(q)\right]n_{-\vec q}
\\
\notag
&&=\sum\limits_{\vec p,\vec q,i}n_{ion}V_0(q)(i\vec q\cdot\vec u_{\vec q})\overline{\psi}_{\vec p +\vec
q,i}\psi_{\vec p,i},
\\
&&H_{e-ph-imp}^{LFR}=\sum_{\vec q} \left[\left(-\nabla_\alpha\left[u_\alpha U_{imp}(\vec r)\right]\right)_{\vec q}\right]n_{-\vec q}
\\
\notag
&&=\sum\limits_{\vec p,\vec q,i}U_{imp}(\vec p'-\vec p)\left(-i(\vec p'+\vec q-\vec p)\cdot\vec u_{\vec q}\right)\overline{\psi}_{\vec p'+\vec q,i}\psi_{\vec p,i},
\eea
where $H_K$ and $H_U$ stand for kinetic and (unaltered) random potential energies respectively.
Usually, this Hamiltonian is used under assumption of static screening in RPA approximation, when $V_0(q)\rightarrow V(q)=(\sum_i\nu_i)^{-1}$, while electro-neutrality in equilibrium implies that $n_{ion}=\sum_j(n_{j})_{eq}=\sum_j\nu_j (p_Fv_F/d)_j$.

\subsection{Canonical transformation}

As Tsuneto has shown \cite{s_tsuneto}, transformation to a CFR in linear approximation in $\vec u$ is equivalent to a canonical transformation $\hat U$:
\bea
&&\psi\rightarrow\hat U\psi,\quad\psi=\hat U^{-1}\psi'
\\
\nonumber
&&\hat U=\left(1+\frac{1}{2}\left\{u_\alpha,\nabla_\alpha \right\}\right)=\left(1+\frac{1}{2}\du+\vec u\cdot\boldsymbol{\nabla}\right)
\\
&&=\left(1+i\vec u_{\vec q}\cdot(\vec p+\vec q/2)\right).
\eea
This transformation alters the Hamiltonian , generating new terms linear in displacement $\vec u$ from $H_K$, $H_U$ and $H_{e-e}$. $H_{e-ph}^{LFR}$ obviously remains unaltered as it is already linear in $\vec u$. However, there is also a special term that arises from the left hand side of Schroedinger equation or, equivalently, from the time derivative in action.

\subsubsection{Time derivative in LHS of Schroedinger equation}

Time derivative in LHS of Schroedinger equation (primes are omitted for brevity) generates the term
\bea
&&i\overline{\psi}\partial_t\psi\rightarrow
\\
\nonumber
&& i\left(1-\frac{1}{2}\du-i\vec u\cdot\vec \partial\right)\overline{\psi}\partial_t\left(1-\frac{1}{2}\du-i\vec u\cdot\vec \partial\right)\psi
\\
\nonumber
&&
=i\overline{\psi}\partial_t\psi-i\overline{\psi}\left(\frac{1}{2}{\rm div}\dot{\vec u}+i\dot{\vec u}\cdot\vec \partial\right)\psi
\\
&&H_{e-ph,LHS}^{CFR}=\sum_{\vec p,\vec q i}\left(\dot{\vec u}_{\vec q} \cdot\vec p\right)\overline{\psi}_{\vec p+\vec q,i}\psi_{\vec p-\vec q,i}
\\
&&
\nonumber
=
\sum_{\vec p,\varepsilon,\omega,\vec q, i}\left(-i\omega\vec u_{\omega,\vec q}\cdot\vec p\right)\overline{\psi}_{\vec p+\vec q/2,\varepsilon+\omega,i}\psi_{\vec p-\vec q/2,\varepsilon,i}
\eea

\subsubsection{Kinetic energy}

\bea
&&
H_{e-ph,1}^{CFR}=H_K\left[\overline{\left(\hat U^{-1}\psi\right)},\hat U^{-1}\psi\right]-H_K\left[\overline{\psi},\psi\right]
\\
\notag
&&=-\sum_{\vec p,\vec q, i}\overline{\psi}_{i}\left[\left(\xi_i(\vec p+\vec q/2)-\xi_i(\vec p-\vec q/2)\right)i\vec p\cdot \vec u\right]\psi_{i}
\\
&&=-\sum_{\vec p,\vec q, i}\overline{\psi}_{\vec p+\vec q/2,i}(i\vec v_{i}\cdot\vec q)(\vec p\cdot \vec u_{\vec q})\psi_{\vec p-\vec q/2,i}
\eea

\subsubsection{Random potential}

\bea
&&H_{e-ph,2}^{CFR}=H_U\left[\overline{\left(\hat U^{-1}\psi\right)},\hat U^{-1}\psi\right]-H_U\left[\overline{\psi},\psi\right]
\\
\notag
&&=-\sum_{\vec p,\vec p',\vec q i}U_{imp}(\vec p'-\vec p)\left[i\vec u_{\vec q}\cdot(\vec p-\vec p')\right] \overline{\psi}_{\vec p'+\vec q/2,i}\psi_{\vec p-\vec q/2,i}
\eea

\subsubsection{Electron-electron interactions}

Electron-electron interaction term is convenient to be analyzed in real space. Electron density is transformed under canonical transformation as
\bea
\notag
&&n_{i}(\vec r)=\overline{\psi}_{i}(\vec r)\psi_{i}(\vec r)\rightarrow
\overline{\left(\hat U^{-1}\psi\right)}_i(\vec r)\left(\hat U^{-1}\psi\right)_i(\vec r),
\\
&&n_{i}(\vec r)\rightarrow n_{i}(\vec r)+\partial_\alpha(u_\alpha(\vec r) n_{i}(\vec r)).
\eea
The contribution to electron-phonon interaction itself is
\bea
\notag
&&H_{e-ph,3}^{CFR}=H_{e-e}\left[\overline{\left(\hat U^{-1}\psi\right)},\hat U^{-1}\psi\right]-H_{e-e}\left[\overline{\psi},\psi\right]
\\
&&=-\sum\limits_{\vec r,\vec r'}u_\alpha(\vec r)\partial_{\alpha} V(\vec r-\vec r')n(\vec r)n(\vec r')
\\\notag
&&=\sum\limits_{\vec p,\vec p',\vec Q, \vec q,i,j}iu_{\vec q,\alpha}Q_\alpha V_0(\vec Q)\overline{\psi}_{\vec p +\vec Q+\vec
q,i}\overline{\psi}_{\vec p',j}\psi_{\vec p'+\vec Q,j}\psi_{\vec p,i}.
\eea

\subsection{Full expression for e-ph interaction in CFR}

Electron-phonon interaction in CFR after canonical transformation is the sum of all parts
\bea
&&H_{e-ph}^{CFR}=H_{e-ph}^{LFR}+H_{e-ph,LHS}^{CFR}
+\sum_iH_{e-ph,i}^{CFR}.
\eea
Two major facts should be emphasized. 

First, electron-phonon-impurity interaction almost cancels out by the part arising from random potential energy:
\bea
&&H_{e-ph-imp}^{LFR}+H_{e-ph,2}^{CFR}
\\
\notag
&&=-\sum\limits_{\vec p,\vec q,i}U_{imp}(\vec p'-\vec p)\left(i\vec q\cdot\vec u_{\vec q}\right)\overline{\psi}_{\vec p'+\vec q,i}\psi_{\vec p,i}
\eea
This cancellation reflects the fact that canonical transformation returns impurities to equilibrium position, while the remaining part is present due to non-uniformity of the transformation.

Second, in Hartree-Fock approximation electron-phonon-ion term is canceled out by the one arising from electron-electron interactions \cite{s_schmid}:
\bea
&&\left(H_{e-ph,3}^{CFR}\right)_{HF}=
\\
\notag
&&\sum\limits_{\vec p,\vec p',\vec Q, \vec q,i,j}iu_{\vec q,\alpha}Q_\alpha V_0(\vec Q)\left\langle\overline{\psi}_{\vec p +\vec Q+\vec
q,i}\psi_{\vec p,i}\right\rangle\overline{\psi}_{\vec p',j}\psi_{\vec p'+\vec Q,j}
\\\notag
&&=-\sum\limits_{\vec p,\vec p' \vec q,j}iu_{\vec q,\alpha}q_\alpha V_0(q)(n_{el})_{eq}\overline{\psi}_{\vec p',j}\psi_{\vec p'-\vec q,j}
\\
&&=-H_{e-ph-ion}^{LFR}
\eea

Thus, electron-phonon interaction in CFR is of the form
\bea
\notag
&&\left(H_{e-ph}^{CFR}\right)_{HF}=-\sum_{\vec p,\vec q, i}\left(\dot{\vec u}_{\vec q}\cdot\vec p\right)\overline{\psi}_{\vec p+\vec q/2,i}\psi_{\vec p-\vec q/2,i}
\\
\label{seq:H_CFR}
&&-\sum_{\vec p,\vec q, i}\overline{\psi}_{\vec p+\vec q/2,i}(i\vec v_{i}\cdot\vec q)(\vec p\cdot \vec u_{\vec q})\psi_{\vec p-\vec q/2,i}
\\
\notag
&&-\sum\limits_{\vec p,\vec q,i}U_{imp}(\vec p'-\vec p)\left(i\vec q\cdot\vec u_{\vec q}\right)\overline{\psi}_{\vec p'+\vec q,i}\psi_{\vec p,i}
\eea
For the problem in question, effects arising from the first term are small in adiabatic parameter $s/v_F$, while possible effects arising from the third one are small in inverse conductance $(p_Fl)^{-1}$. Thus, they may be safely omitted.

Finally, there is screening by electron-electron interactions $H_{e-e}$. In this paper we start with bare electron-phonon vertices $H_{e-ph}^{CFR}$ and take into account screening explicitly (for example, we sum up diagrammatic ladders given on the Fig.\ref{sfig:vertices}). 
In a number of papers  a bit different formulation of the problem is used;
namely, it is assumed from the very beginning  that full screening of 
Coulomb interaction takes place.
Within such an approach  electron-electron interaction does not appear in the  explicit  form in the calculations;  instead, an additional term that describes the
above  screening is added into the electron-phonon vertex.
In Eq.(36) below we present, for the sake of completeness, the corresponding form of the electron-phonon interaction  in the  co-moving frame (with first and third  terms of (S35) omitted)
\bea
&&\left(H_{e-ph}^{CFR}\right)_{scr}=
\\
\notag
&&-\sum_{\vec p,\vec q, i}\overline{\psi}_{\vec p+\vec q/2,i}(i\vec v_{i}\cdot\vec q)(\vec p\cdot \vec u_{\vec q})\psi_{\vec p-\vec q/2,i}
\\
\notag
&&+\sum_{\vec p,\vec q, i}\overline{\psi}_{\vec p+\vec q/2,i}\left(V_{RPA}\sum\limits_j\nu_j\left(\frac{p_Fv_F}{d}\right)_j \du\right)\psi_{\vec p-\vec q/2,i},
\eea
where we assume static screening in RPA approximation, $V_{RPA}=(\sum_j\nu_j)^{-1}$.

\section{II. Derivation of the phonon kinetic equation}

Here we derive the phonon quantum kinetic equation via the Keldysh
diagrammatic technique (see Ref.\onlinecite{s_Kamenev-Levchenko} for
details).
 For simplicity, we consider only the longitudinal phonons;
 the transverse phonons may be analyzed in a similar way. The Green's functions are matrices
 in the  Keldysh space:
\bea
&&\hat D=\left(
\begin{array}{cc}
D^K & D^R
\\
D^A & 0
\end{array}
\right),
\\
\notag
&&\hat D_0^{-1}=\left(
\begin{array}{cc}
0 & (D_0^A)^{-1}
\\
(D_0^R)^{-1}    & 0
\end{array}
\right),\,
\hat \Sigma=\left(
\begin{array}{cc}
0 & \Sigma^A
\\
\Sigma^R    & \Sigma^K
\end{array}
\right),
\eea
where $D_0$ and $D$ are the bare and exact phonon Green's functions,
respectively and $\Sigma$ is  the self-energy part. The Keldysh
component $(\dots)^K$ may be parametrized as
\be\label{seq:G-Kel} D^K=D^R\circ F-F\circ D^A \ee
where $F$ is related to the phonon distribution function; in
equilibrium  $F=\coth(\omega/2T)$. The sign $\circ$ means the
convolution in the time domain.
 A bare retarded(advanced) phonon Green's function is
\be \label{seq:D_0}
D_0^{R(A)}(\omega,q)=\frac{1}{\hbar^2\rho_m}\frac{1}{(\omega\pm
i0)^2-v_{s}^2q^2}, \ee where $\rho_{m}$ is the mass density.
To derive the kinetic equation one  employs the Dyson equation for
the matrix Green functions  which together with
Eq.(\ref{seq:G-Kel})results in:
\bea  &&(\hat D_0^{-1}-\hat \Sigma)\hat D=1,\quad\Rightarrow
\\
\Rightarrow &&\left[D_0^{-1},F\right]=\Sigma^K-\Sigma^R\circ
F-F\circ\Sigma^A.\label{seq:matrix-kinur} \eea
According to  Eq.(\ref{seq:D_0}) $D_{0}^{-1}$ contains the second
derivative with respect to the corresponding time argument of
$F\equiv F(t_{1},t_{2})$. Then the commutator in
Eq.(\ref{seq:matrix-kinur}) reduces to the first derivative
$\partial_{t}$ with respect to the "slow" combination
$t=t_{1}+t_{2}$ multiplied by the Fourier transform $-i\omega$ of
the derivative with respect to the "fast" combination $t_{1}-t_{2}$.
Averaging over the whole volume of the sample, one finds \be
2i\rho_m\omega\partial_tF=\Sigma^K-\Sigma^R\circ F-F\circ\Sigma^A.
\ee In this equation we assume $F(t_{1},t_{2})\rightarrow
F(\omega;t)$. We also omit the terms describing the external source
that pumps energy into electronic system.

A common action of the source term and the e-ph energy relaxation
leads to a stationary energy distribution of both electrons and
phonons. If the e-ph energy relaxation is much slower than the
electron-electron one, a quasi-equilibrium situation with two
temperatures $T_{el},\,T_{ph}$ is realized \cite{s_comment}.
 In this case the phonon distribution function is $F(\omega,T_{ph})=\coth(\omega/2T_{ph})$
 while $\Sigma^K$ is a quasi-equilibrium quantity.  If in addition a weak e-ph interaction is
 assumed, electron-phonon interaction enters only trivially as the
 square of the e-ph matrix element in
  the Keldysh self-energy part $\Sigma^{K}$. In this approximation $\Sigma^{K}$ depends
only on the \emph{electron}
 temperature, $\Sigma^K=F(\omega,T_{el})(\Sigma^R-\Sigma^A)$,
\bea
\notag
&&2i\rho_m\omega\partial_tF(\omega,T_{ph}(t))=\left[F \cdot (\Sigma^R-\Sigma^A)\right](\omega,q,T_{el})
\\
\label{seq:kin_eq_F}
&&-\left[F \cdot (\Sigma^R-\Sigma^A)\right](\omega,q,T_{ph}).
\eea
Since the phonon decay rate is relatively low,
$\tau_{ph}^{-1}\ll\omega$, phonons are well-defined quasiparticles
and the quasiparticle distribution function is sharply peaked around
the phonon "mass-shell" $\omega=sq$. Thus, the quantities entering
in the  R.H.S. of Eq.(\ref{seq:kin_eq_F}) should be taken at
$\omega=sq$, see Ref.~\cite{s_Kamenev-Levchenko}.
For convenience we introduce a standard  phonon distribution function $B (\omega) =(1/2)(F(\omega) -1)$ and define a decay rate $\tau_{ph}^{-1}$:
\bea
\label{seq:kinetic1}
\tau_{ph}^{-1}(\omega,T_{el})=\frac{1}{\rho_m\omega}\Im\Sigma^R(\omega,q,T_{el})|_{\omega=sq},
\\
\partial_tB(\omega,T_{ph}(t))=\frac{B(\omega,T_{el})-B(\omega,T_{ph})}{\tau_{ph}(\omega,T_{el})}.
\label{seq:kinetic2}
\eea
In order to obtain an electron-phonon heat flow, we have to multiply Eq.(\ref{seq:kinetic2}) by both phonon density of states
 $\nu_{ph}(\omega)=\omega^{2}/2\pi^2v_{s}^3$ and by energy, and integrate
over $\omega$:
\bea \label{seq:J1} &&\mathcal{J}=\int\limits_0^\infty
d\omega\,\nu_{ph}(\omega)\,\omega\,\partial_t\,B(\omega,T_{ph}(t))
\\
\notag
&&=\int\limits_0^{\infty}d\omega\,\nu_{ph}(\omega)\,\frac{\omega}{\tau_{ph}(\omega,T_{el})}
\left[B(\omega,T_{el})-B(\omega,T_{ph})\right] \eea
The incoming and outgoing energy flows may be defined as follows:
\bea
\label{seq:J2}
&&\mathcal{J}(T_{el},T_{ph})=J_+(T_{el})-J_-(T_{ph},T_{el}),
\\
\notag
&&J_+(T_{el})=\int\limits_0^{\infty}d\omega\,\nu_{ph}(\omega)\,\frac{\omega}{\tau_{ph}(\omega,T_{el})}B(\omega,T_{el}),
\\
\notag
&&J_-(T_{ph},T_{el})=\int\limits_0^{\infty}d\omega\,\nu_{ph}(\omega)\,\frac{\omega}{\tau_{ph}(\omega,T_{el})}B(\omega,T_{ph}),
\eea
In general, $J_-$ depends on both $T_{el}$ and $T_{ph}$ due to the
effect of electron-electron interactions on the decay rate
$\tau_{ph}^{-1}$, see Eq.(\ref{seq:kinetic1}). However such an effect is
absent within the RPA approximation for charge screening 
which we use
here. Thus our result for the phonon lifetime does not depend on
temperature explicitly, $\tau_{ph}=\tau_{ph}(\omega)$.
 In this case the expressions for heat flows are simplified and $J_-$ becomes a function of the phonon temperature $T_{ph}$ only.

The electronic temperature relaxation rate $\tau_E^{-1}$ can be
obtained from the expression for the heat flow $\mathcal{J}_{e-ph}$.
Each phonon branch contributes as
\bea
\label{seq:tau_E}
 \tau_{E,\alpha}^{-1}=\frac{1}{C_e}\left.\frac{\partial \mathcal{J}_{e-ph,\alpha}}{\partial T_{el}}\right|_{T_{ph}=T_{el}}
\\ \nonumber
=\frac{1}{4C_eT^2}\int\limits_0^{\infty}d\omega
\frac{\omega^2
\nu_{ph}(\omega)}{\sinh^2(\omega/2T)}\tau_{ph,\alpha}^{-1}(\omega,T)
\eea
where $C_e \propto \nu T $ is the electronic
specific heat.  The full rate is then
$\tau_{E,f}^{-1}=\tau_{E,l}^{-1}+(d_{ph}-1)\tau_{E,tr}^{-1}$, where
$(d_{ph}-1)$ is the number of  transverse phonon polarizations.
Electrons are characterized by the quasi-equilibrium distribution function if their intrinsic inelastic scattering
time $\tau_{ee}$ is much shorter
than the temperature relaxation time $\tau_E$ due to interaction with phonons.

Finally, let us mention that in 2D electron systems the phonon decay
rate $\tau_{ph}(\vec q)$ depends on the angle $\theta$ between the
phonon momentum $\vec q$ and the direction normal to the plane of
the 2DEG. Thus, in the equations like
(\ref{seq:J1},\ref{seq:J2},\ref{seq:tau_E}) an angle-averaged decay rate should
be used: \be \left<\tau_{ph}^{-1}\right>_\theta(\omega)=\frac12
\int_0^{\pi}(\sin\theta d\theta)\,\tau^{-1}_{ph}(\omega,\theta) \ee

\section{III. Diagrammatic derivation }

\begin{figure}[h]
\includegraphics[width=1.0\linewidth]{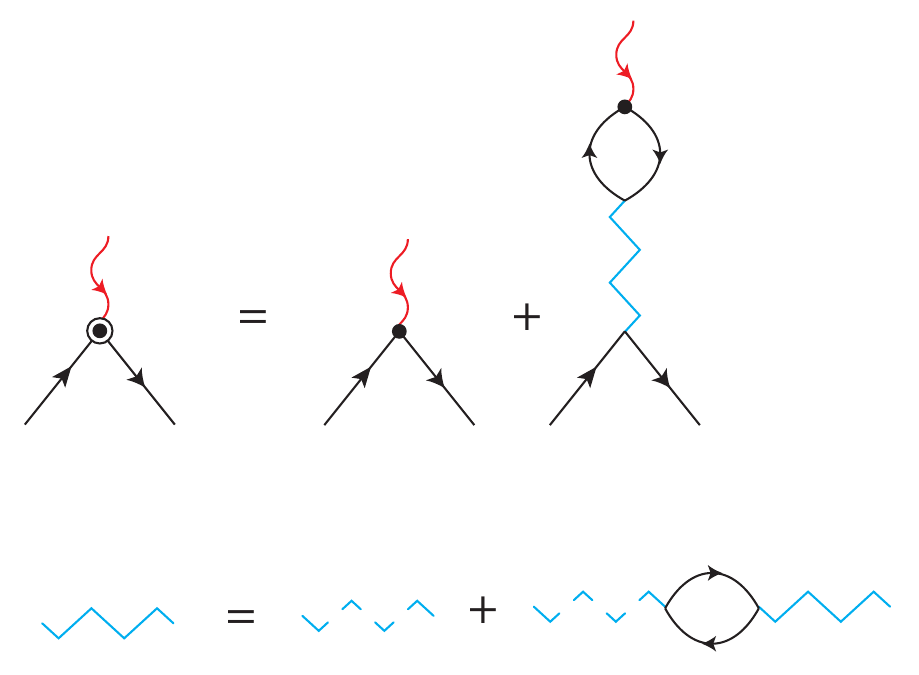}
\caption{Bare and screened e-ph vertices. Black, curvy red and dashed blue zig-zag lines are electron,
phonon and Coulomb interaction propagators respectively. Bold zig-zag lines stand for statically screened
Coulomb interaction (with screening by empty electron bubbles only). }
\label{sfig:vertices}
\end{figure}

We present here the derivation of some of our results in the standard diagrammatic form,
as it was done in the most papers on this subject~\cite{s_schmid,s_reyzer,s_kravtsov_yudson}.

The bare electron-phonon vertex corresponding to the Hamiltonian (\ref{seq:H}) is of
tensor structure in the space of electron species:
\be
\label{seq:gamma}
\hat{\Gamma}_{bare}=
\left(
\begin{array}{ccc}
\Gamma_1 & &
\\
    & \ddots &
    \\
    &   & \Gamma_N
\end{array}
\right).
\ee
The diagonal structure of (\ref{seq:gamma}) corresponds to our
assumption about the absence of inter-branch mixing. Now one should
take into account static Coulomb screening, which generates scalar
counter-term
\be
\label{seq:gamma_c}
\hat{\Gamma}_C=-\frac{\sum_i \Gamma_i\nu_i}{V_0^{-1}(q)+\sum_i\nu_i}\hat{1}
\ee
The full screened vertex is  then a sum $\hat{\Gamma}_{full}=\hat{\Gamma}_{bare}+\hat{\Gamma}_{C}$.
The structure of these vertices is presented in Fig.\ref{sfig:vertices}.
The deformation potentials $\Gamma_i$ averaged over FS are usually  approximated as \cite{s_schmid}
\be
\Gamma_i=\left[\left.\left(\Gamma_{bs}(\vec p)\right)\right|_{FS}-p_Fv_F/d_e\right]_i
\ee
where $\Gamma_{bs}$ represents the
lattice-induced deformation potential
under the lattice strain and $p_Fv_F/d_e$ represents the averaged
electron liquid stress tensor. We will consider the simplest model
where the
lattice contribution is uniform in momentum space
$\Gamma_{bs}(\vec p)=\Gamma_{bs}$ and thus is reduced to the shift of electron band\, (see however
Ref.~\cite{s_comment2}).
\begin{figure}[h]
\includegraphics[width=1.0\linewidth]{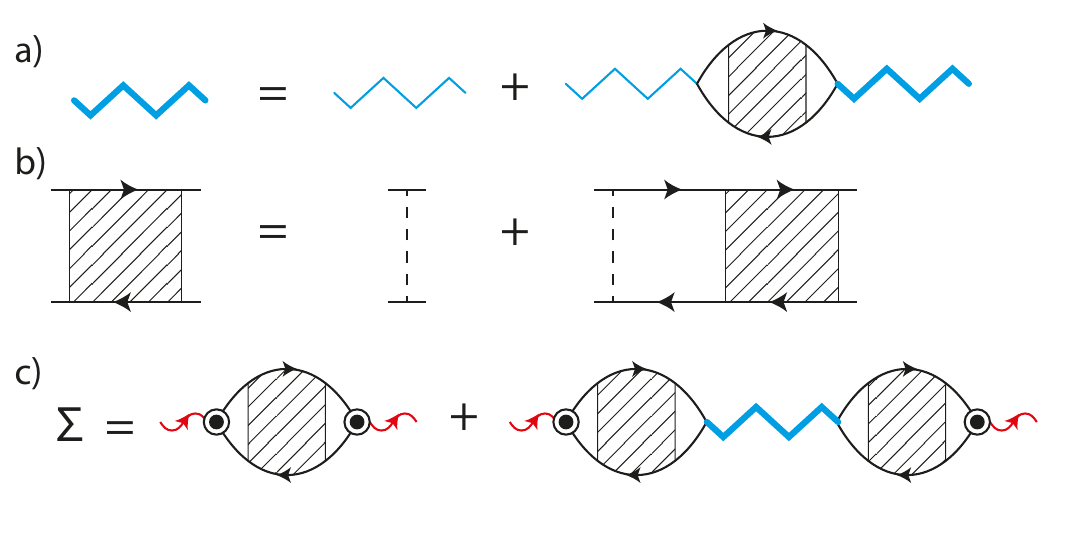}
\caption{a) Dynamically screened Coulomb interaction, where diffusion is taken into account.
Namely, an impurity ladder is summed up
\ %
b) The impurity ladder. Here black dashed line represents impurity correlator
$\left\langle U(\vec r)U(\vec r')\right\rangle$.
c) Diagrams contributing to phonon self energy. This figure encompasses the very
general case with arbitrary number of electron types and arbitrary Coulomb interaction
strength. Inside the electron bubbles summation goes over all electron branches
}
\label{sfig:2}
\end{figure}

In order to evaluate the decay rate and the electron cooling rate we
need to calculate the imaginary part of the phonon self energy
$\Sigma$. It is represented by the diagrams shown in  a
Fig.\ref{sfig:2}.
 The second diagram is important when the screening is essentially dynamic.
 Below we demonstrate  few particular examples  how the diagrammatic description works.

\subsection{A. Imperfect screening}

In this Subsection we consider the case of identical spectral branches, but assuming now that
screening is incomplete and electron density variations are allowed.
Then the full e-ph vertex given by the sum of Eq.(\ref{seq:gamma}) and
Eq.(\ref{seq:gamma_c}) is diagonal:
\be
\hat\Gamma_{s}=\frac{\Gamma}{1+N_f\nu V_0(q)} \ee
where $N_f$ is a number of identical electron branches. Typically it
is equal to $N_f=2N_v$ where $N_v$ is the number of identical
valleys in a semiconductor. For example, $N_v=6$ for a bulk silicon
or $N_v=2$ for graphene. The first diagram from the Fig.\ref{sfig:2}c
thus gives
\be
\Sigma_1= \left(\frac{\Gamma}{1+\nu N_f V_0(q)}\right)^2q^2N_f\nu\frac{i\omega}{-i\omega+Dq^2}
\ee
 The second diagram turns out to be crucial for a dynamical screening regime, when $\omega\geq Dq^2$:
\bea
&&\Sigma_2=\left(\frac{\Gamma q}{1+\nu N_f V_0(q)}\right)^2\left(N_f\nu\frac{i\omega}{-i\omega+Dq^2}\right)^2
\\
\notag
&&\times \left(-\frac{1}{V_0^{-1}(q)+N_f\nu Dq^2/(-i\omega+Dq^2)}\right)
\eea
Summing up these contributions we obtain $\Sigma=\Sigma_1+\Sigma_2$:
\bea
&&\Sigma=
\frac{\Gamma^2q^2}{1+N_f\nu V_0(q)}\times
\\
\nonumber
&&\times N_f\nu\left(\frac{i\omega(i\omega +Dq^2+(N_f\nu V_0(q)Dq^2)}
{\omega^2 +(Dq^2)^2(1+N_f\nu V_0(q))^2}\right).
\eea
The phonon decay rate is determined by the imaginary part of $\Sigma$:
\be
\tau_{ph,ML}^{-1}=
\frac{\Gamma^2}{\rho_m}\frac{N_f\nu Dq^2}{v_{s}^2 +(Dq)^2(1+N_f\nu
V_0(q))^2},
\\
\label{seq:F_C} \ee and the corresponding enhancement factor is
\be
\mathcal{F}_{C}(q)= 1+
\frac{1}{c_l}\frac{(\Gamma/p_Fv_F)^2}{(v_{s}/v_F)^2+d_e^{-2}(q^2l^2)(1+N_f\nu
V_0(q))^2} \label{seq:Fc}
\ee
These results coincide with the ones in the main text for $N_f=2$. Thus, we have shown equivalence of the
approach using macroscopic kinetic equation and diagrammatic technique.

\subsection{B. Several spectral branches}

Now we switch to the situation of complete screening
of Coulomb interaction, so the electro-neutrality condition is
obeyed exactly. In such a case the difference in the coupling
constants $\Gamma_i$ corresponding to different electron branches is
crucial. We consider here the simplest case of two spectral
branches.
Then the full screened e-ph vertex given by the sum of
Eqs.(\ref{seq:gamma}) and (\ref{seq:gamma_c}) is traceless:
\be
\hat{\Gamma}_{s}= \left(
\begin{array}{cc}
\frac{\Gamma_1-\Gamma_2}{2} &
    \\  & -\frac{\Gamma_1-\Gamma_2}{2}
\end{array}
\right).
\ee
The diagrams shown in  Fig.\ref{sfig:2}c give
\be
\Sigma= 2\times\left(\frac{\Gamma_1-\Gamma_2}{2}\right)^2q^2\nu\frac{i\omega}{-i\omega+Dq^2}
\ee
 Thus, the phonon decay rate due to the Mandelstam-Leontovich mechanism is
\bea
&&\tau_{ph}^{-1}=\frac{1}{\rho_m\omega}\left[\Im\Sigma(\omega,q)\right]_{\omega=sq}
\\
\nonumber &&=\frac{(\Gamma_1-\Gamma_2)^2}{2\rho_m}\frac{\nu
Dq^2}{v_{s}^2+(Dq)^2}, \eea
and the total decay rate is enhanced (with respect to the classical
Pippard result) by the factor
\be
\label{seq:F_MB}
\mathcal{F}(q)=1+\frac{1}{4c_l}\left(\frac{\Gamma_1-\Gamma_2}{p_Fv_F}\right)^2\frac{v_F^2}{v_{s}^2+(Dq)^2}
\ee
If the asymmetry in spectral branches arise due to the Zeeman
splitting, then $\Gamma_1-\Gamma_2=(2/d_e)g\mu_B H$: \be
\label{seq:F_H}
\mathcal{F}_{H}(q,h)=1+\frac{1}{d_e^2c_l}\frac{v_F^2h^2}{v_{s}^2+(Dq)^2}
\ee where $h=g\mu_B H/2\varepsilon_F \ll 1$ is dimensionless magnetic
field (we neglect here $h$-dependencies of other parameters, like
difference $D_1\neq D_2$, which is negligible  at $h\ll 1$).

Note that the effect is absent in zero magnetic field, as the
time-reversal symmetry does not allow  spin density fluctuations to be excited
by phonons.

\subsection{C. Spin-orbit coupling and  interbranch scattering}

In this Subsection we revisit the case of the two inequivalent
branches of electron spectrum and consider the (previously
neglected) role of the interbranch scattering, using the
Zeeman-splitting as an example. Qualitatively, the spin-flip
scattering with a rate $\tau_r^{-1}$ leads to non-conservation of
the total spin and thus limits the magnitude of any effect which is
related to slow spin diffusion. Formally it is described by the
modification of the spin diffusion propagator:
\be
\frac{Dq^2}{-i\omega + Dq^2} \Rightarrow \frac{Dq^2}{-i\omega + Dq^2+1/2\tau_r}
\ee
which leads to the replacement of  Eq.(\ref{seq:F_H}) by
\be
\mathcal{F}_{H}(q,h)=1+\frac{1}{d_e^2c_l}\frac{v_F^2q^2h^2}{\omega^2+(Dq^2+1/2\tau_{r})^2}.
\label{seq:F_H_r}
\ee

Below we calculate $\tau_r$ for the  case of 2D electron gas with
the Zeeman splitting  induced by magnetic field applied in the 2DEG
plane in the $x$-axis direction.
\begin{figure}[h]
\includegraphics[width=0.5\linewidth]{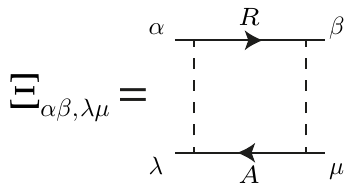}
\caption{Diffuson self energy $\hat{\Xi}$, $\mathcal{\hat D}=(1/2\pi\nu\tau)(1-\hat{\Xi})^{-1}$.
Capital letters $A$ and $R$ stand for advanced and retarded electron Green functions respectively.
}
\label{sfig:diffuson_se}
\end{figure}
We assume a relatively weak spin-orbit (SO) interaction leading to the spin-orbit band splitting $\Delta_{SO}\ll\Delta_H=g\mu_B H$.
For definiteness we consider the Rashba-type SO coupling with the
spin-dependent part of the Hamiltonian being equal to
\be
H_{H+SO}=-\frac{\Delta_H}{2}\hat
\sigma_x+\frac{\Delta_{SO}}{2}(\sigma_xn_y-\sigma_yn_x),
\ee
where
$\vec n=\vec p/|\vec p|$ is a unit vector in the direction of
momentum. The elastic scattering time turns out to be equal for both
quasiparticle branches(in the absence of electron-hole asymmetry):
\bea &&\hat G^R(\varepsilon,\vec p)=\sum\limits_{\pm}\frac{\hat
P_\pm}{\varepsilon-\xi\mp\Delta(\vec n)/2+i/2\tau},
\\
&&\hat P_\pm=\frac{1}{2}\left(1\pm\frac{\Delta_H\sigma_x-\Delta_{SO}(\sigma_xn_y-\sigma_yn_x)}{\Delta(\vec n)}\right),
\\
&&\Delta(\vec n)=\sqrt{\Delta_H^2+\Delta_{SO}^2-2\Delta_H\Delta_{SO}n_x}.
\eea
where $\hat G^R$ is the retarded electron Green's function. In order
to find the relaxation rate, we evaluate the diffuson self energy
for zero frequency and momentum ($\omega=0,\,\vec q=0$),
Fig.\ref{sfig:diffuson_se}:
\be \hat\Xi=\frac{1}{2\pi\nu\tau}\int\frac{d\vec p}{(2\pi)^2}\hat
G^A(0,\vec p)\otimes \hat G^R(0,\vec p) \ee A simple calculation in
a manner similar to that of Ref.\onlinecite{s_skvortsov} leads to the
following result for the diffuson self energy at $q=\omega=0$:
\begin{widetext}
\be
\hat\Xi=\left(1-\hat S_x^2\left[1-\frac{1}{1+\tau^2\Delta_H^2}\right]
-\frac{i\tau\Delta_H}{1+\tau^2\Delta_H^2}\hat S_x\right)
-\frac{\Delta_{SO}^2}{2\Delta_H^2}\left(\frac{\tau^2\Delta_{H}^2}{1+\tau^2\Delta_H^2}\hat S_y^2-\tau^2
\Delta_{H}^2\frac{(3+\tau^4\Delta_H^4)}{(1+\tau^{2}\Delta_H^2)^3}\hat S_x^2-\frac{4i\tau^3\Delta_H^3}
{(1+\tau^2\Delta_H^2)^3}\hat S_x\right)
\label{eq:Xi}
\ee
\end{widetext}
with $\vec{\hat {S}}=(\hat
1\otimes\boldsymbol{\hat{\sigma}}-\boldsymbol{\hat{\sigma}}\otimes\hat1)/2$
being the total spin of electron-hole pair. We are interested in the
$S_x=0$ subspace only as it hosts two eigenvalues of our interest.
Naturally, the singlet mode($S=0$) corresponding to the charge
density propagation remains unaffected, $\Xi_{S=0}=1$,
 while the triplet mode($S=1,S_x=0$) representing a spin density diffusion does decay:
\be \Xi_{
S=1,S_x=0}=1-\frac{1}{2}\frac{\Delta_{SO}^2}{\Delta_H^2+\tau^{-2}}
\label{seq:SigmaS1} \ee leading to the following result for the spin
decay rate 
\be 
\tau_r^{-1}\equiv
\tau_{so}^{-1}=\frac{\Delta_{SO}^2}{\Delta_H^2+\tau^{-2}}\tau^{-1}\ll\tau^{-1}
\label{seq:tau_r} 
\ee
In the course of derivation of Eq.(\ref{seq:SigmaS1})
we used an identity $\langle S=1,S_x=0|\hat S^2_y|S=1,S_x=0\rangle=1
$. We emphasize that Eq.(\ref{seq:tau_r}) was derived for weak SO
interaction, $\Delta_{SO}\ll\Delta_H$.

\section{IV. Angular dependence of ultrasonic attenuation}

We start here by the quasi-2D case  when the thickness of a
semiconductor film is much larger than the Fermi wavelength but
still smaller than the phonon wavelength, $\lambda_F\ll b\ll
\lambda_{ph}$. In this case electron diffusion is two-dimensional
and only the component of phonon momentum parallel to the plane
enters into the diffusion propagator where a replacement $
Dq^2/(-i\omega + Dq^2) \Rightarrow D{\bf q}_{\parallel}^2/(-i\omega
+ D{\bf q}_{\parallel}^2)$ should be made.  The result is that
Eqs.(\ref{seq:F_H},\ref{seq:Fc}) should be replaced by
\bea
\label{seq:F_Q2D1}
&&\mathcal{F}_{C}(q) =1 +
\frac{c_l^{-1}(\Gamma/p_Fv_F)^2\sin^2\theta}{(v_{s}/v_F)^2+(2/d_e)^2g_\Box^2(e^2/\varepsilon\hbar
v_F)^2},
\\
\label{seq:F_Q2D2}
&&\mathcal{F}_H(q)=1+\frac{1}{4c_l}\frac{v_F^2h^2\sin^2\theta}{v_{s}^2+(Dq)^2\sin^4\theta},
\eea
where the last equation is given for relatively strong screening $\nu V_0(q)\gg1$.

The true 2D case, however, should be discussed specially.  While the
result for the case of imperfect screening at $|\Gamma|\gg p_{F}v_{F}$
is identical to Eq.(\ref{seq:F_Q2D1}), the magnetic-field induced
effect (arising from the momentum-dependent part of the
electron-phonon vertex) may  behave differently.

For a sufficiently thin film electron motion in the direction
perpendicular to the plane is fully quantized, thus the expression
for electron-phonon vertex becomes 
\bea \label{seq:Gamma_2D}
&&\Gamma_{1,2}=\Gamma_{bs}\du - \frac{1}{m}\langle p_z^2\rangle_z
(iq_zu_z)
\\
\notag &&-  \frac{p^{2}_{\parallel}}{2m}(iq_xu_x+iq_yu_y)
\\
\notag &&=\left(\Gamma_{bs} - \frac{1}{m}\langle
p_z^2\rangle_z\cos^2\theta -
\frac{p^{2}_{\parallel}}{2m}\sin^2\theta\right) \du, 
\eea
where 
\be
\frac{p^{2}_{\parallel}}{2m}=\varepsilon_{F}\pm g\mu_B H
-\frac{\langle p^{2}_{z}\rangle_{z}}{2m}
\ee
with $\pm$ corresponding
to the spin-up and spin-down electrons.
\begin{figure}[h]
\includegraphics[width=1.0\linewidth]{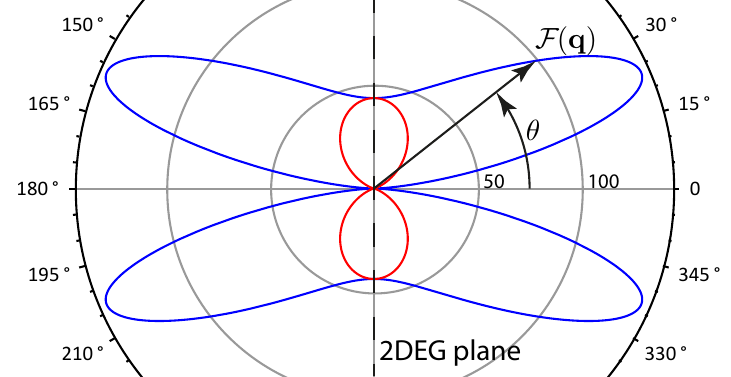}
\caption{Angular dependence of ultrasonic attenuation for magnetic-field-controlled case, Eq.\ref{seq:F_H_2D}.
Red and blue curves represent ($\gamma_\parallel=0,\gamma_\perp=0$) and ($\gamma_\parallel=0,\gamma_\perp=1$)
respectively. Both plots are given for parameters of InSb sample from the main text at frequency $\omega=200\,\mathrm{MHz}$}
\label{sfig:angular_dependence}
\end{figure}
Here we have taken an average $\langle\cdot\rangle_z$ over the
ground state corresponding to the motion in the perpendicular
direction and $p_{\parallel}$ is the momentum of an in-plane motion.
We assume that the matrix element $\langle p^{2}_{z}\rangle_{z}$
does not depend on the spin degree of freedom as well as on the
electron density. Thereby we  disregard any possible orbital effects
of magnetic field and consider the Zeeman interaction only which
results in two different  momenta $p_{\parallel,\uparrow}$ and
$p_{\parallel,\downarrow}$ corresponding to the in-plane motion for
the up and the down spin projections. The momentum-independent
component of the vertex does not contribute to the
magnetic-field-controlled relaxation under the condition of perfect
screening (according to Eq.(\ref{seq:gamma_c}) it is screened out
completely). Eqs.(\ref{seq:F_MB}) explicitly implies that only
asymmetric part of the vertex contributes:
\be
\Gamma_1-\Gamma_2=g\mu_B H\sin^2\theta \du
\ee

However, in real 2D electron systems, such as heterostructures, the
lattice strain also affects momentum-dependent part of the quasiparticle spectrum. In other words, it alters electron effective mass. We limit ourselves to the particular case when
\bea
&&\delta m^{-1}=-m^{-1}(\gamma_\perp\cos^2\theta+\gamma_\parallel\sin^2\theta)\du,
\\ \nonumber
&& \Gamma_1-\Gamma_2=
g\mu_B H\times[(1+\gamma_\parallel)\sin^2\theta+\gamma_\perp\cos^2\theta] \du.
\eea
Thus, the magnetic-field-controlled enhancement becomes equal to
\bea
\label{seq:F_H_2D}
&&\mathcal{F}_{H}(q,h)=1+
\\
\notag
&&\frac{v_F^2h^2\sin^2\theta}{v_{s}^2+(Dq)^2\sin^4\theta}[(1+\gamma_\parallel)\sin^2\theta+\gamma_\perp\cos^2\theta]^2
\eea The results (\ref{seq:F_Q2D1},\ref{seq:F_Q2D2},\ref{seq:F_H_2D})
show the angular dependence of ultrasonic attenuation which may
exhibit a characteristic cross-like pattern exemplified in
Fig.\ref{sfig:angular_dependence}.

There are two additional issues which should be addressed to make
the above analysis really quantitative: (i) in a general case the
incident longitudinal(transverse) acoustic wave reflected off the
free surface produces both longitudinal and transverse reflected
waves, (ii) in the true 2D case diffusion modes could be generated
by transverse phonons as well.
 However, these effects do not seem to lead
to any qualitative change of our results and we will postpone the
corresponding studies for the future.

\section{V. Electron-phonon heat flow}

In this Subsection we use previously obtained results for the phonon decay rate, Eqs.(\ref{seq:F_Q2D1}-\ref{seq:F_Q2D2}),
to derive an expression for
the electron-phonon heat flow in a true 2D electron gas structure.
We start by the spin density diffusion effects. At the lowest
temperatures $T\ll T_{H}^{(1)}=\hbar v_{s}^2/D$, the enhancement does
not depend on temperature and angle, being just a numerical factor:
\be \label{seq:J_H1} J_{H} = g_\Box\frac{\pi^3 p_F^2v_F^2h^2}{126\hbar^6\rho_m
v_{s}^7}A_1 T^6, \ee where
$A_1=(48(1+\gamma_\parallel)^2+16(1+\gamma_\parallel)\gamma_\perp+6\gamma_\perp^2)/105$
and $g_\Box = k_F l$ is the dimensionless conductance of the 2DEG.
 For higher temperatures, $T\gg T_{H}^{(1)}$, the enhancement factor behaves as
 $\mathcal{F}\propto T^{-2}$. However, the resulting expression for the e-ph heat flow depends significantly
  on the angular structure of the vertex
\bea
\nonumber
&&J_{H} =  \frac{1}{g_\Box}\frac{\pi p_F^4
h^2}{120\hbar^6 \rho_m v_{s}^3}\left(A_2+\gamma_\perp^2\ln
T/T_{H}^{(1)}\right)T^4
\\
\label{seq:J_H2}
&&A_2=\frac{4}{3}(1+\gamma_\parallel)(1+\gamma_\parallel+\gamma_\perp)
\\
\nonumber
&&+\gamma_\perp^2\left(\frac{90}{\pi^4}\zeta^\prime(4)-\frac{5+6\gamma}{6}\right)
\eea
where $\zeta^\prime$ is the derivative of Riemann zeta function and $\gamma=0.577$ is the Euler's constant respectively.
If the normal strain does not alter electron mass ($\gamma_\perp
=0$) the angular dependence $\mathcal{F}(q,\theta)$ does not lead to
 dominance of small $\theta$ in the corresponding integral.
Then the heat flow is a pure power-law $J\propto T^4$. Otherwise,
the shape of $\mathcal{F}(q,\theta)$ is rather peculiar (see
Fig.\ref{sfig:angular_dependence}) and an additional $\ln T$ factor
appears due to the contribution of small angles
$\theta_{H}\sim\sqrt{T_{H}^{(1)}/T}$. Note that the $T^{4}\ln T$
behavior is slower than $T^{6}$ and at temperatures
\be
T>T_{H}^{(2)}=T_{H}^{(1)}\,\sqrt{1+\frac{h^{2}v_{F}^{2}}{v_{s}^{2}\overline{{\cal
F}}_{C}}},\;\;\;\quad T_{H}^{(1)}=\hbar\frac{v_{s}^{2}}{D}
\label{seq:T2H}
\ee
the effect of spin
fluctuations is smaller than the $H$-independent contribution
proportional to $T^{6}$.
In the above  Eq.(\ref{seq:T2H}) we denote as ${\overline{\cal F}}_{C}$  the
$H$-independent enhancement factor Eq.(39) averaged over angles $\theta$.
The spin-orbit interaction suppresses the effect of spin
fluctuations at low temperatures:
$$
T<T_{H}^{(SO)}=\frac{\hbar v_{s}}{\sqrt{D\tau_{SO}}},
$$
so that the condition for this effect to be observed is
$$
T_{H}^{(SO)}<T_{H}^{(2)}.
$$
The temperature dependence of the e-ph heat flow in the most
favorable case $T_{H}^{(SO)}<T_{H}^{(1)}$ is shown in
Fig.\ref{sfig:cooling}.
\begin{figure}[h]
\includegraphics[width=1.0\linewidth]{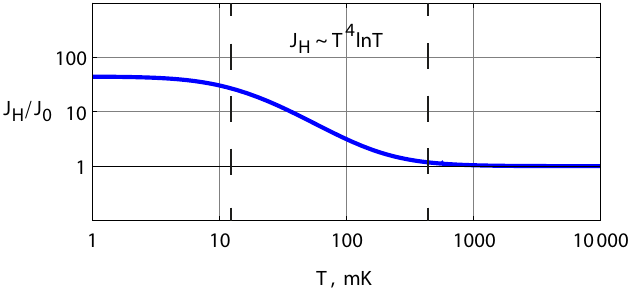}
\caption{Temperature dependence of electron-phonon heat flow
enhanced by the spin-diffusion. Plotted is the ratio of the
spin-diffusion part $J_{H}(T)$ and the local part Eq.(51). In the
wide temperature region $T_{H}^{(1)}<T<T_{H}^{(2)}$, where
$T_{H}^{(1)}\sim 20\,\mathrm{mK}$, $T_{H}^{(2)}\sim 500\,\mathrm{mK}$, the outgoing
heat flow $J_{H}(T)\propto T^{4}\,\ln(T/T_{H}^{(1)})$. The
parameters of the 2D system were: electron density
$n=10^{11}\,\mathrm{cm^{-2}}$, electron effective mass $m=0.1\,m_{0}$,
($m_{0}$ is the free electron mass), the Fermi momentum
$1/k_{F}=12\, \mathrm{nm}$, the Fermi energy $E_{F}=2.4\,\mathrm{meV}$, dimensionless conductance $p_Fl=10$, the Lande
$g$-factor $g=5$, the spin-orbit band splitting
$\Delta_{R}=0.002\,E_{F}$ (corresponding to the Dresselhaus
splitting in InP at the corresponding density); the anisotropy
parameters in Eq.(47) are $\gamma_{\parallel}=0$, $\gamma_{\bot}=1$.
The magnetic field $H=8\mathrm{T}$ corresponds to $h=0.5$. }
\label{sfig:cooling}
\end{figure}
For the effects of charge diffusion on the e-ph heat flow in the 2D
electron system with no additional screening of
interactions($\varepsilon=const$) the situation is similar to that
of Eq.(\ref{seq:J_H1}):
\be
\label{seq:J_C1}
 J_{C_{2D}}
=\frac{1}{g_\Box}\frac{2\pi^3p_F^2}{63\hbar^6\rho_m v_{s}^5}
\left(\frac{\varepsilon
\hbar v_F}{e^2}\right)^2\left(\frac{\Gamma}{p_Fv_F}\right)^2T^6, \ee
where the effective dielectric constant is an arithmetic mean of the
dielectric constants of the media on both sides of the 2D system:
$\varepsilon=(\varepsilon_{1}+\varepsilon_{2})/2$.
 The
enhancement is thus reduced to a temperature-independent  factor.

The result is the same for the  geometry with additional screening
by a metallic gate at temperatures $T\gg \hbar v_{s}/b$ with $b$ being the
distance between the 2D electron plane and the gate. At lower
temperature $T\ll \hbar v_{s}/b$ the effective dielectric permittivity is $q$-dependent: $\varepsilon(q)=\varepsilon(\coth
qb+1)/2\approx \varepsilon/2qb$ which transforms Coulomb interaction
into a short-range one,  $V_0(q)=4\pi e^2 b$. The behavior becomes
similar to Eqs.(\ref{seq:J_H2})  however, with the crossover
temperature $T_{H}^{(1)}$ replaced by  $T_{C}^{(1)}=(1/g_\Box
k_Fb)(v_{s}/v_F)^2(\varepsilon \hbar v_F/e^2)E_F$: \bea \label{seq:J_Cg1}
&&J_{C_{2D+gate}}=\frac{1}{g_\Box}\frac{\pi p_F^4}{240\hbar^6\rho_m
v_{s}^3}
\\
\nonumber &&\times \frac{1}{k_F^2b^2}\left(\frac{\varepsilon\hbar
v_F}{e^2}\right)^2\left(\frac{\Gamma}{p_Fv_F}\right)^2T^4\ln\frac{T}{T_{C}^{(1)}}
,\, \qquad T\gg T_{C}^{(1)},
\\
\label{seq:J_Cg2} &&J_{C_{2D+gate}}=g_\Box\frac{2\pi^3\Gamma^2}{63\hbar^6\rho_m
v_{s}^7}T^6,\, \qquad  T\ll T_{C}^{(1)}, \eea
Similarly to (\ref{seq:J_H2}), Eq.(\ref{seq:J_Cg1}) contains $\ln T$
coming from the angles $\theta_C\sim\sqrt{T_{C}^{(1)}/T}$.

We also note that Eq.(\ref{seq:J_Cg1}),(\ref{seq:J_Cg2}) represent
only the Mandelstam-Leontovich contribution due to charge diffusion.
At high enough temperatures $$T>T_{C}^{(2)}\sim
T_{C}^{(1)}\,\left(\frac{v_{F}}{v_{s}}\right)$$ this contribution is
smaller than the $J^{(0)}_{tr}$ arising from local processes corresponding to the
transverse phonon Pippard's ultrasound attenuation, Eq.(2) in the
paper:
\bea
\label{seq:J0}
J_{0}=g_\Box\frac{2\pi^3p_F^2}{189\hbar^6\rho_m
v_{s,t}^5}T^6
\eea
Thus for temperatures $T>T_{C}^{(2)}$ the $T^{6}$ law is
restored.

We should note that in real heterostructures the deformation
potential $\Gamma$ is in general anisotropic.However, this fact does
not lead to any profound changes like suppression of logarithmic
behavior $\propto\ln T$ in Eq.(\ref{seq:J_Cg1}). This would require a
highly anisotropic deformation potential $\Gamma\propto\sin^2\theta$
which we do not expect.

\section{VI. A particular example: thin film of $\mathrm{InSb}$.}

Here we consider the enhancement of ultrasound attenuation in an
$\mathrm{InSb}$ thin film  with electron density
$n=10^{11}\mathrm{cm^{-2}}$ and thickness $d \geq 10 - 20 \mathrm{nm}$ on
the  $\mathrm{SiO_2}$ substrate. We consider the spin effect
of parallel magnetic field applied to the film, so its thickness $d$
is chosen to be relatively small (slightly larger than Fermi
wavelength) in order to avoid orbital effects of magnetic field. In spite of the fact that piezoelectric coupling is present in InSb, it is irrelevant for in-plane longitudinal phonons \cite{s_piezo} as matrix element of piezoelectric interaction for such phonons is equal to zero.
Thus, we assume that phonon wavevector $\vec q$ is parallel to the 2DEG plane. 

Effective mass of $\mathrm{InSb}$ is equal to $\,m=0.014m_0$ \cite{s_ioffe},
spin-orbital band splitting is $\Delta_{SO}\approx0.11\mathrm{meV}$
\cite{s_insb_so1,s_insb_so2}  and the electron mean-free-path is
supposed to be relatively long,  $p_Fl=50$. At such parameters the
Fermi energy $\varepsilon_{F}=0.02\, \mathrm{eV}$, while the deformation
potential $\Gamma=-13.12\, \mathrm{eV}$ \cite{s_dp_insb}. A $\mathrm{SiO_2}$ substrate is
characterized by $v_{s}=6\times10^5\mathrm{cm/s}$ and the effective
dielectric constant $\varepsilon=(1+3.9)/2\approx2.5$. Finally, we
use Eqs.(\ref{seq:F_H_r}) and (\ref{seq:tau_r})  with the specified
sample/material parameters. The resulting plots are given in the
Fig.1 of the main text.

\section{VII. A general expression for phonon decay rate: a phenomenological derivation}

Here we derive a general expression for ultrasonic attenuation using
a phenomenological approach of diffusive electron transport. We
start by the system of equations
\bea
\label{seq:equation_motion}
\left\{
\begin{aligned}
&\partial_t n^{(i)}+\mathrm{div} \vec j^{(i)}=0,
\\
&\vec j^{(i)} =-D^{(i)}\vec \nabla n^{(i)} + \kappa^{(i)} \vec
F^{(i)}
\\
&U^{(i)}= \int {\cal V}_0(\vec r-\vec r')\sum\limits_{j}\delta n^{(j)}(\vec r') + \Gamma^{(i)}\du
\ea
\right.
\eea
Fourier transforming the set we get
\be
\left\{
\begin{aligned}
&(-i\omega + D^{(i)}q^2)n^{(i)}=-\kappa^{(i)}q^2U^{(i)}
\\
&U^{(i)}= V_0(q)\sum\limits_{j}n^{(j)}(\vec q) + (i\vec q\cdot\vec u)\Gamma^{(i)}
\ea
\right.
\ee
Due to Coulomb interaction the solution for i-th branch depends on the dynamics of total density. Thus, the solution is
\be
n^{(i)}=-\Pi^{(i)}(\omega,q)\left(\Phi^{(i)}-\Phi_C(\omega,q)\right)
\ee
where $\Pi^{(i)}(\omega,q)=\kappa^{(i)}q^2/(-i\omega+D^{(i)}q^2)$ is
a response function, $\Phi^{(i)}=(i\vec q\cdot\vec u)\Gamma^{(i)}$,
\be \label{seq:Phi}
\Phi^C=\frac{V_0(q)\sum_{i}\Pi^{(i)}\Phi^{(i)}}{1+V_0(q)\sum_{i}\Pi^{(i)}}.
\ee
Here $\Phi^C$ describes dynamic Coulomb counteraction.
 To obtain the phonon decay rate we have to evaluate the dissipation power in a unit volume
\begin{widetext}
 \bea
 &Q_t=\frac{1}{2}\sum\limits_i\Re\left(\vec j^{(i)}\cdot \vec F^{*(i)}\right)
=\frac{1}{2}\sum\limits_i\Re\left[-i\omega(\Phi^{(i)}-\Phi_C(\omega,q))\Pi^{(i)}(\Phi^{(i)}-\Phi_C(\omega,q))^*\right]
 \\
 \notag
 &
=\frac{\omega}{2}\sum\limits_i\Im\left[(\Phi^{(i)}-\Phi_C(\omega,q))\Pi^{(i)}(\Phi^{(i)}-\Phi_C(\omega,q))^*\right].
 \eea
\end{widetext}
 and the energy of acoustic wave:
\be
 E_w=\frac{\rho_m}{2}\omega^2u_m^2
\ee Finally, for the phonon decay rate we get
\bea
\tau_{ph}^{-1}=\frac{Q_t}{E_w}=\frac{q^2}{\rho_m\omega}\sum\limits_i\Im\left[\Gamma^{(i)}_{s}\Pi^{(i)}(\Gamma^{(i)}_s)^*\right]
\\
=\frac{q^2}{\rho_m\omega}\sum\limits_i\Gamma^{(i)}_{s}\Im\left[\Pi^{(i)}\right](\Gamma^{(i)}_s)^*
\eea
 where $\Gamma_s=(\Phi^{(i)}-\Phi_C(\omega,q))/\du$  can  be considered as a dynamically screened vertex.

 \subsection{A. Multiple branches}

 We analyze here the case of two quasiparticle branches and very strong bare Coulomb potential, $V(q)\rightarrow\infty$.
 The Coulomb counteraction term takes the form
 \be
 \Phi^C=\frac{\sum_{i}\Pi^{(i)}\Phi^{(i)}}{\sum_{i}\Pi^{(i)}}
 \ee
 that exactly fixes total density $\delta(n_1+n_2)=0$. Thus  the phonon decay rate becomes equal to
 \begin{widetext}
 \be
 \tau_{ph}^{-1}=\frac{q^2}{\rho_m\omega}\Im\left[\frac{(\Gamma_1-\Gamma_2)\Pi_2}{\Pi_1+\Pi_2}\Pi_1\frac{(\Gamma_1-\Gamma_2)
 \Pi_2^*}{(\Pi_1+\Pi_2)^*}\right.
 +\left.\frac{(\Gamma_2-\Gamma_1)\Pi_1}{\Pi_1+\Pi_2}\Pi_2\frac{(\Gamma_2-\Gamma_1)\Pi_1^*}{(\Pi_1+\Pi_2)^*}\right]
 =\frac{(\Gamma_1-\Gamma_2)^2q^2}{\rho_m\omega}\Im\left[\Pi_1^{-1}+\Pi_2^{-1}\right]^{-1}
 \ee
 \end{widetext}
 Using also the fact that $\kappa_i=\nu_i D_i$ we obtain:
\be
\tau_{ph}^{-1}=\frac{(\Gamma_1-\Gamma_2)^2}{\rho_m}\frac{\nu_*D_*q^2}{v_{s}^2+(D_*q)^2}
\label{seq:tau_MB} \ee
where $v_{s}=\omega/q$ is the sound velocity and
$\nu_*=(\nu_1^{-1}+\nu_2^{-1})^{-1}$,
$D_*=\nu_*^{-1}((\nu_1D_1)^{-1}+(\nu_2D_2)^{-1})^{-1}$ are the
effective density of states and diffusion coefficient respectively.
To obtain the total ultrasonic attenuation we also have to take into
account the PIC result:
\be
(\tau_{ph}^{-1})^{(0)}=c_l\frac{p_F^2}{\rho_m}(\nu_1D_1+\nu_2D_2)q^2
\ee
This equation coincides with Eq.(1) of the main text if both
electron branches are identical $\nu_1=\nu_2,\, D_1=D_2$. Thus, for
the total attenuation rate
 we obtain:
\be
\tau_{ph}^{-1}=c_l\frac{p_F^2}{\rho_m}(\nu_1D_1+\nu_2D_2)q^2+\frac{(\Gamma_1-\Gamma_2)^2}{\rho_m}\frac{\nu_*D_*q^2}{v_{s}^2+(D_*q)^2}
\ee
and the  enhancement factor $\mathcal{F}=\tau_{ph}^{-1}/(\tau_{ph}^{-1})^{(0)}$ is
 \be
 \mathcal{F}_{MB}(q)=1+\frac{1}{c_l}\left(\frac{\Gamma_1-\Gamma_2}{p_Fv_F}\right)^2\frac{\nu_*D_*}
 {\nu_1D_1+\nu_2D_2}\frac{v_F^2}{v_{s}^2+(D_*q)^2}
 \ee
An important particular case is that of the Zeeman splitting by an
in-plane magnetic field $H$ for a 2D electron system ($d_e=2$). Here
an asymmetry appears between the spin-up and spin-down electron
branches (while $\nu_1=\nu_2=\nu$):
$\varepsilon_{\uparrow(\downarrow)}=\varepsilon_F\pm\mu H/2$,
$D_{\uparrow(\downarrow)}=D(1\pm\mu H/2\varepsilon_F)$. However, the
most important asymmetry is the one in the vertex $\Gamma$ that
arises from the momentum-dependent part of the electron-phonon
coupling:
\be
\Gamma_\uparrow-\Gamma_\downarrow=(\partial\Gamma/\partial\varepsilon_F)\mu H
\ee
In the simplest model, where the only effect of the strain upon  electron spectrum is its overall shift,
the density-dependent contribution arises from the  stress $\langle p_\alpha v_\beta\rangle_{FS}$ of electron liquid only:
 \be
 \Gamma(p_F)=\Gamma_0-p_Fv_F/2\rightarrow\partial\Gamma/\partial\varepsilon_F=-1
 \ee
Introducing the dimensionless magnetic field $h=\mu
H/2\varepsilon_F$, we arrive at the result
  \be
  \mathcal{F}_{H}(q,h)=1+\frac{v_F^2h^2(1-h^2)}{v_{s}^2+(Dq(1-h^2))^2}
  \ee
Finally we discus the effect of inter-branch scattering. In fact, it
modifies the response function
 \be
\frac{D_*q^2}{-i\omega+D_*q^2}\Rightarrow\frac{D_*q^2}{-i\omega+D_*q^2+1/2\tau_{so}}
 \ee
where $\tau_{SO}$ is the characteristic time of inter-branch mixing
so labeled in analogy with the spin-orbit mixing of electron
branches with different spin projections. Tt limits the diffusion
enhancement factor
\be
\frac{D_*q^2}{\omega^2+(D_*q^2)^2}\Rightarrow\frac{D_*q^2}{\omega^2+(D_*q^2+1/2\tau_{so})^2}
\ee
and the final result  becomes equal to
\be \label{seq:FH} \mathcal{F}_H(q,h) =
1+\frac{q^2v_F^2h^2(1-h^2)}{(q v_{s})^2 +
(Dq^2(1-h^2)+1/2\tau_{so})^2} \ee

\subsection{B. Imperfect screening}

 Another case of interest is the case of two quasiparticle branches with identical parameters
 $\Gamma_{1(2)}=\Gamma,\,\nu_{1(2)}=\nu,\,D_{1(2)}=D$ but finite strength of
 Coulomb interaction. In this case no asymmetry is present and thus asymmetric electron modes cannot be excited.
 However, finite Coulomb interaction and incomplete screening allows
 density fluctuations which diffusive relaxation leads to the
 enhancement of ultrasound attenuation and the e-ph energy flow:
 \begin{widetext}
 \be
  \tau_{ph}^{-1}=\frac{q^2}{\rho_m\omega}\Im\left[2\times\left(\Gamma-\frac{2V_0\Pi\Gamma}{1+2V_0\Pi}\right)
  \Pi\left(\Gamma-\frac{2V_0\Pi\Gamma}{1+2V_0\Pi}\right)^*\right]
  =\frac{q^2}{\rho_m\omega}\Im\left[\frac{2\Gamma^2\Pi}{|1+2V_0\Pi|^2}\right]
  =\frac{\Gamma^2}{\rho_m}\frac{2\nu Dq^2}{v_{s}^2+(2\nu V_0+1)^2(Dq)^2}
 \ee
 \end{widetext}
 \be
 \mathcal{F}_{C}= 1 + \frac{d_e^2}{c_l}\frac{(\Gamma/p_Fv_F)^2}{d_e^2(v_{s}/v_F)^2+(q^2l^2)(1+2\nu V_0(q))^2}
 \ee

 For a 2D geometry and  Coulomb interaction $V_0(q)=2\pi e^2/\varepsilon q$ which is still relatively strong, $\nu V\gg1$,
 $\nu V(q)ql\gg s/v_F$, the result acquires a simple, frequency-independent form:
\be \mathcal{F}_{C_{2D}}= 1 +
\frac{1}{c_lg_\Box^2}\left(\frac{\varepsilon\hbar
v_F}{e^2}\right)^2\left(\frac{\Gamma}{p_Fv_F}\right)^2
\label{seq:F_C_Supp} \ee where $g_\Box$ is dimensionless conductance per
square in $e^2/h$ units. The expression is valid also for a quasi-2D
sample, when the phonon wavelength $\lambda$ is much larger than the
width of quasi-2D system $d_s$.

In the most general case,  the bare Coulomb potential $V_0(q)$
acting between conduction electrons can be written in terms of some
dispersive dielectric response $\varepsilon(q)$, as $V_0(q)=2\pi
e^2/q \varepsilon (q)$.  Therefore Eq.(\ref{seq:F_C_Supp}) can be used
in order to extract $\varepsilon(q)$ dependence from the measured
phonon relaxation rate.

An important special case is presented by a 2D electron gas with a
metal gate placed nearby which additionally screens
electron-electron interaction. Here $\varepsilon(q) =
\varepsilon/(1-e^{-2qb})$ and $V_0(q)=(2\pi e^2/\varepsilon
q)(1-e^{-2qb})$, where $b$ is the distance between electron plane
and  gate parallel to it. As long as phonon wavelength is shorter
than distance $b$, the presence of the gate may be ignored, while
for short wavelengths, when $qb\ll1$, Coulomb interaction becomes
effectively-short-range, $V_0(q)=4\pi e^2 b/\varepsilon$. Exactly
for this region of low frequencies (temperatures) enhancement factor
$\mathcal{F}$ is
\be \mathcal{F}_{C_{2D+gate}}= 1 + \frac{4
(\Gamma/p_Fv_F)^2}{(v_{s}/v_F)^2+(q^2l^2)(4\pi\nu
e^2b/\varepsilon)^2} \ee
We see that the crossover between $\mathcal{F}\propto\omega^{-2}$
and $\mathcal{F}\propto\omega^0$ behavior emerges at
\be \nonumber \hbar\omega_{cross}\sim
\frac{1}{g_\Box}\frac{1}{k_Fb}\left(\frac{v_{s}}{v_F}\right)^2\left(\frac{\varepsilon\hbar
v_F}{e^2}\right)E_F. \ee This equation can be used for an
experimental determination of the background dielectric constant
$\varepsilon$.


\begin{thebibliography}{99}
\bibitem{nu} In the case of 2D electron gas we consider an array of 2D electron planes with the distance $L$ between them. Then in Eq.(1) $\nu\rightarrow\nu_{2d}/L$.
\end{thebibliography}
\begin{thebibliography}{99}
\bibitem{s_comment} We assume  that phonon distribution function is an equilibrium one, either due to phonon-phonon inelastic scattering, or
due to the role of external phonon bath.
\bibitem{s_comment2}
Piezoelectric coupling may be incorporated in our scheme, in this case vertex will depend on phonon momentum and one will have to replace
constant $\Gamma$ by the  $\vec q$-dependent function.
\end{thebibliography}
\end{document}